\documentclass[prd,twocolumn,showpacs,nofootinbib,floatfix]{revtex4}

\usepackage{graphicx}
\usepackage{amsmath,amsfonts}
\usepackage{dcolumn}

\newcommand{\td}[1]{{\rm d}#1} 
\newcommand{\Metric}{g} 
\newcommand{\Snormal}{n} 
\newcommand{\Lapse}{\alpha} 
\newcommand{\CLapse}{\tilde\alpha} 
\newcommand{\Shift}{\beta} 
\newcommand{\SMetric}{\gamma} 
\newcommand{\CMetric}{\tilde\gamma} 
\newcommand{\CF}{\psi} 
\newcommand{\ExCurv}{K} 
\newcommand{\TrExCurv}{K} 
\newcommand{\TFExCurv}{A} 
\newcommand{\CTFExCurv}{\tilde{A}} 
\newcommand{\SRicci}{\bar{R}} 
\newcommand{\SRicciS}{\bar{R}} 
\newcommand{\CRicciS}{\tilde{R}} 
\newcommand{\dtime}{\partial_t} 
\newcommand{\SCD}{{\bar\nabla\!}} 
\newcommand{\CCD}{{\tilde\nabla}\!} 
\newcommand{\LieD}[1]{{{\cal L}_{#1}}}
\newcommand{\CLD}[1]{(\tilde{\mathbb L}{#1})} 

\newcommand{\CMtd}{\tilde{u}} 
\newcommand{\Bnormal}{s} 
\newcommand{\CBnormal}{\tilde{s}} 
\newcommand{\BMetric}{h} 
\newcommand{\CBMetric}{\tilde{h}} 
\newcommand{\BExCurv}{H} 
\newcommand{\PBExCurv}{J} 
\newcommand{\ONull}{k} 
\newcommand{\INull}{\acute{k}} 
\newcommand{\ONExCurv}{\Sigma} 
\newcommand{\INExCurv}{\acute\Sigma} 
\newcommand{\Oexpansion}{\theta} 
\newcommand{\Iexpansion}{\acute\theta} 
\newcommand{\Oshear}{\sigma} 
\newcommand{\Ishear}{\acute\sigma} 
\newcommand{\BCD}{D} 
\newcommand{\CBCD}{\tilde{D}} 
\newcommand{\perpShift}{{\Shift_{\!\mbox{\tiny$\perp$}}}} 
\newcommand{\parShift}{{\Shift_{\mbox{\tiny$\parallel$}}}} 
\newcommand{\FMetric}{f} 

\begin{document}

\title{Excision boundary conditions for black hole initial data}

\author{Gregory B. Cook}\email{cookgb@wfu.edu}
\affiliation{Department of Physics, Wake Forest University,
		 Winston-Salem, North Carolina\ \ 27109}

\author{Harald P. Pfeiffer}\email{harald@tapir.caltech.edu}
\affiliation{Theoretical Astrophysics, California Institute of Technology,
 		 Pasadena, California\ \ 91125}

\date{\today}

\begin{abstract}
We define and extensively test a set of boundary conditions that can
be applied at black hole excision surfaces when the Hamiltonian and
momentum constraints of general relativity are solved within the
conformal thin-sandwich formalism.  These boundary conditions have
been designed to result in black holes that are in quasiequilibrium
and are completely general in the sense that they can be applied with
any conformal three-geometry and slicing condition.  Furthermore, we
show that they retain precisely the freedom to specify an arbitrary
spin on each black hole.  Interestingly, we have been unable to find a
boundary condition on the lapse that can be derived from a
quasiequilibrium condition.  Rather, we find evidence that the lapse
boundary condition is part of the initial temporal gauge choice.  To
test these boundary conditions, we have extensively explored the case
of a single black hole and the case of a binary system of equal-mass
black holes, including the computation of quasi-circular orbits and
the determination of the inner-most stable circular orbit.  Our tests
show that the boundary conditions work well.
\end{abstract}


\maketitle

\section{Introduction}
\label{sec:introduction}

The simulation of black-hole systems necessarily starts with the
specification of initial data.  In order for such simulations to yield
astrophysically relevant results, the initial data must be constructed
to be astrophysically realistic.  Achieving this is the goal of
efforts being made to improve black-hole, and in particular black-hole
binary, initial data.  It has become clear that {\em all} of the freely
specifiable pieces of the initial data, including the boundary
conditions, must be chosen carefully to respect the physical content
of any system we wish to simulate.  In this paper we will focus on the
boundary conditions that are required when a black hole's interior is
excised from the initial-data domain.

In Ref.~\cite{Cook-2002}, one of the authors proposed a set of
boundary conditions that were intended to yield a black hole that was
in quasiequilibrium.  These conditions were chosen to be consistent with
the desire to create a binary system fully in quasiequilibrium.  It is
reasonable to expect such a system will be astrophysically realistic
if the black holes in the binary are sufficiently far apart and in a
nearly circular orbit.  In this paper, we refine and extensively test
these boundary conditions in the cases of a single black hole and
a pair of equal-mass black holes in a binary system.

The most significant refinement of the quasiequilibrium boundary
conditions over the original version in Ref.~\cite{Cook-2002} is to
the procedure for specifying the spin of each black hole.  The
analysis below shows that the spin must be chosen in a very specific
way in order to be compatible with the assumptions of
quasiequilibrium.  Fortunately, the procedure still allows for a
completely arbitrary specification of the spin and this procedure is
compatible with {\em any} choice of the conformal three-geometry.

A significant result from our tests on the original set of
quasiequilibrium boundary conditions is that the proposed lapse
boundary condition is not viable.  We will show below that this
boundary condition is {\em degenerate} when combined with the other
quasiequilibrium boundary conditions.  Furthermore, the nature of this
degeneracy can be easily understood.  We conjecture that the boundary
condition on the lapse is not fixed by quasiequilibrium considerations
but is, rather, a part of the initial temporal gauge choice.  Below,
we provide analytical and numerical evidence to support this
conjecture.

For a single black hole, the quasiequilibrium boundary conditions
allow for the construction of initial data that yield true stationary
spacetimes.  Doing so, however, requires that {\em all} of the freely
specifiable data be chosen in a way that is compatible with a
stationary black hole.  In particular, it requires that the conformal
three-geometry be chosen correctly.  Unfortunately, there is still no
general prescription for choosing an appropriate conformal
three-geometry.  Because of this, we have chosen to perform all of our
tests on a flat conformal three-geometry.  By testing the boundary
conditions for the three cases of a single static, spinning, and
boosted black hole, we will be able to test both the quasiequilibrium
boundary conditions and the effect that the assumption of conformal
flatness has on the resulting initial data.  For equal-mass black hole
binaries, we will extensively test the special cases of corotating and
irrotational black holes.  The numerical results we obtain will be
compared with post-Newtonian results and previous numerical results
for both cases.

We begin in Sec.~\ref{sec:conf-thin-sandw} with a review of the
conformal thin-sandwich decomposition of the constraints, and then
derive the quasiequilibrium boundary conditions in
Sec.~\ref{sec:quasi-equil-bound}.  In
Secs.~\ref{sec:quas-solut-single} and \ref{sec:quasi-circ-orbits} we
apply the boundary conditions to the cases of a single black hole and
to equal-mass black hole binary systems.  Finally, in
Sec.~\ref{sec:discussion} we further explore the effectiveness of
these boundary conditions.

\section{The conformal thin-sandwich decomposition}
\label{sec:conf-thin-sandw}

In this work, we will use the standard \mbox{3+1} decomposition with the interval written as
\begin{equation}
\td{s}^2 = -\Lapse^2 \td{t}^2 
  + \SMetric_{ij}(\td{x}^i +\Shift^i\td{t})(\td{x}^j +\Shift^j\td{t}),
\label{eq:3+1_interval}
\end{equation}
where $\SMetric_{ij}$ is the 3-metric induced on a $t=\mbox{const.}$
spatial hypersurface, $\Lapse$ is the lapse function, and $\Shift^i$
is the shift vector.  The extrinsic curvature of the spatial slice,
$\ExCurv_{ij}$, is defined by
\begin{equation}
\ExCurv_{\mu\nu} \equiv -\frac12\SMetric^\delta_\mu\SMetric^\rho_\nu
          \LieD{\Snormal}\SMetric_{\delta\rho},
\label{eq:ExCurv_def}
\end{equation}
where $\LieD{\Snormal}$ denotes the Lie derivative along the unit
normal to the spatial slice, $\Snormal^\mu$.  Einstein's equations, in
vacuum, then reduce to four sets of equations.  Two are evolution
equations for the spatial metric and extrinsic curvature:
\begin{equation}
\dtime\SMetric_{ij} = -2\Lapse\ExCurv_{ij} + 2\SCD_{(i}\Shift_{j)},
\label{eq:metric_evol}
\end{equation}
and
\begin{eqnarray}
\dtime\ExCurv_{ij} &=& -\SCD_i\SCD_j\Lapse 
+ \Lapse\left[\SRicci_{ij} 
- 2\ExCurv_{i\ell}\ExCurv^\ell_j 
+ \TrExCurv\ExCurv_{ij}\right]
\nonumber \\ && \mbox{}
+\Shift^\ell\SCD_\ell\ExCurv_{ij} 
+ 2\ExCurv_{\ell(i}\SCD_{j)}\Shift^\ell.
\label{eq:ExCurv_evol}
\end{eqnarray}
The remaining two are the constraint equations
\begin{equation}
\SRicciS + \TrExCurv^2 - \ExCurv_{ij}\ExCurv^{ij} = 0
\label{eq:Hamiltonian_const}
\end{equation}
and
\begin{equation}
\SCD_j(\ExCurv^{ij} - \SMetric^{ij}\TrExCurv) = 0.
\label{eq:Momentum_const}
\end{equation}
Here, $\SCD_i$, $\SRicci_{ij}$, and $\SRicciS$ are, respectively, the
covariant derivative, Ricci tensor, and Ricci scalar associated with
the spatial metric $\SMetric_{ij}$.  Finally, the trace of the extrinsic
curvature is denoted $\TrExCurv \equiv \ExCurv^i_i$.

The task of constructing initial data for a Cauchy evolution via
Einstein's equations requires that we decompose the constraints in
such a way that we can specify how the constrained, gauge, and
dynamical degrees of freedom are associated with the initial data.  In
this paper, we are primarily interested in initial data associated
with systems in quasiequilibrium.  Because of this, it is natural to
use the the {\em conformal thin-sandwich} decomposition of the
constraints\cite{york-1999,Pfeiffer-York-2003}.  This decomposition is
particularly useful in this situation because quasiequilibrium is a
dynamical concept, and this decomposition retains a close connection
to dynamics that is lost in most other decompositions of the
constraints (cf Refs.~\cite{cook-2000,Cook-2002}).

The conformal thin-sandwich decomposition employs a York--Lichnerowicz
conformal decomposition of the metric and various other
quantities\cite{lichnerowicz-1944,york-1971,york-1972}.  The conformal
factor, $\CF$, is defined via
\begin{equation}
\SMetric_{ij} \equiv \CF^4\CMetric_{ij},
\label{eq:Metric_decomp}
\end{equation}
where $\CMetric_{ij}$ is a ``conformal metric''.  The time derivative
of the conformal metric is introduced by the definitions
\begin{align}
\CMtd_{ij} &\equiv \dtime\CMetric_{ij} \label{eq:Conf_u_def},\\
\CMetric^{ij}\CMtd_{ij} &\equiv 0. \label{eq:Tr_u_def}
\end{align}
From this, it follows that the tracefree extrinsic curvature
$\TFExCurv^{ij}\equiv\TrExCurv^{ij}-\frac13\SMetric^{ij}\TrExCurv$ takes
the form
\begin{equation}
\TFExCurv^{ij} =\frac{\CF^{-10}}{2\CLapse}\left[\CLD{\Shift}^{ij} -
\CMtd^{ij}\right],
\label{eq:ExCurv_decomp}
\end{equation}
where $\CLapse\equiv\CF^{-6}\Lapse$ is the conformal lapse function, and
$\CMtd^{ij}=\CMtd_{kl}\CMetric^{ik}\CMetric^{jl}$. 
Furthermore, $\CLD{V}$ is the conformal-Killing (or longitudinal)
operator acting on a vector, defined by
\begin{equation}
\CLD{V}_{ij} \equiv 2\CCD_{(i}V_{j)} - \mbox{$\frac23$}\CMetric_{ij}\CCD_kV^k,
\label{eq:conf-Killing_op}
\end{equation}
where $\CCD_k$ is the covariant derivative compatible with $\CMetric_{ij}$.

Notice that this decomposition of $\ExCurv_{ij}$ incorporates the
kinematical variables of the \mbox{3+1} decomposition, that is, the
shift vector $\Shift^i$ and the lapse function $\Lapse$ through the
conformal lapse $\CLapse$.  It also includes the trace-free time
derivative of the conformal metric, $\CMtd_{ij}$.  Below, the conformal
tracefree extrinsic curvature will be useful,
\begin{equation}
\CTFExCurv^{ij} \equiv \CF^{10}\TFExCurv^{ij}= \frac{1}{2\CLapse}
\left[\CLD{\Shift}^{ij} - \CMtd^{ij}\right].
\label{eq:CTFExcurv_def}
\end{equation}

Within the conformal thin-sandwich formalism, one must specify
$\CMetric_{ij}$, $\CMtd_{ij}$, $\TrExCurv$, and $\CLapse$.  With these
quantities defined, the Hamiltonian (\ref{eq:Hamiltonian_const}) and
momentum (\ref{eq:Momentum_const}) constraints take the form of a
coupled set of elliptic equations that determine $\CF$ and $\Shift^i$.
In terms of our conformally decomposed variables, the Hamiltonian
constraint (\ref{eq:Hamiltonian_const}) can be written
\begin{equation}
\CCD\,^2\CF - \mbox{$\frac18$}\CF\CRicciS
- \mbox{$\frac1{12}$}\CF^5\TrExCurv^2
+ \mbox{$\frac18$}\CF^{-7}\CTFExCurv_{ij}\CTFExCurv^{ij} = 0,
\label{eq:CTS_Ham_con}
\end{equation}
where $\CRicciS$ is the Ricci scalar associated with $\CMetric_{ij}$,
and the momentum constraint (\ref{eq:Momentum_const}) as
\begin{equation}
\CCD_j\left(\mbox{$\frac{1}{2\CLapse}$}\CLD\Shift^{ij}\right)
-\mbox{$\frac{2}{3}$}\CF^6\CCD^{\;i}\TrExCurv 
       - \CCD_j\left(\mbox{$\frac1{2\CLapse}$}\CMtd^{ij}\right)=0.
\label{eq:CTS_mom_con}
\end{equation}

The freely-specified data includes the conformal metric,
$\CMetric_{ij}$, and its time derivative, $\CMtd_{ij} =
\dtime\CMetric_{ij}$, as well as the trace of the extrinsic curvature,
$\TrExCurv$, and the conformal lapse, $\CLapse$.  It is possible, and
desirable, to make the set of freely-specified data more symmetric by
choosing to specify the time derivative of the extrinsic curvature
instead of the conformal lapse.  This is possible because these two
quantities are related by the trace of Eq.~(\ref{eq:ExCurv_evol}).
The resulting equation is an elliptic equation for the conformal lapse
that is coupled to both the Hamiltonian and momentum constraints,
(\ref{eq:CTS_Ham_con}) and (\ref{eq:CTS_mom_con}).  This equation can
be written as
\begin{eqnarray}
\CCD\,^2(\CF^7\CLapse)
-(\CF^7\CLapse)\left( \mbox{$\frac18$}\CRicciS
+ \mbox{$\frac5{12}$}\CF^{4}\TrExCurv^2 
+ \mbox{$\frac78$}\CF^{-8}\CTFExCurv_{ij}\CTFExCurv^{ij}\right)
\nonumber \\ \mbox{}
=- \CF^5\left(\dtime\TrExCurv-\Shift^k\CCD_k\TrExCurv\right).
\label{eq:Const_TrK_eqn}
\end{eqnarray}

The statement made earlier, that the conformal thin-sandwich
decomposition has a close connection to dynamics is now clear.  Not
only does this decomposition incorporate the kinematical variables of
the \mbox{3+1} splitting, but fully half of the freely-specifiable data
consist of time derivatives of fundamental fields.  In particular, we
are free to choose the conformal metric and the trace of the extrinsic
curvature ($\CMetric_{ij}$ and $\TrExCurv$) {\em and} the time
derivatives of these fields ($\dtime\CMetric_{ij}\equiv\CMtd_{ij}$ and
$\dtime\TrExCurv$).

Remaining to be determined are the conformal factor, $\CF$, the
conformal lapse, $\CLapse$, and the shift vector, $\Shift^i$.  These
are determined by solving Eqs.~(\ref{eq:CTS_Ham_con}),
(\ref{eq:CTS_mom_con}), and (\ref{eq:Const_TrK_eqn}) as a coupled set
of elliptic equations.  Formulating a well-posed system requires that
we impose boundary condition.  Typically, these systems are solved
under the assumption that the spacetime is asymptotically flat.  If we
let $r$ denote a coordinate radius measured from the location of the
center of energy of the system, then as $r\to\infty$ we have that
\begin{subequations}\label{eq:outer-BCs}
\begin{align}
\label{eq:CF_BC_infty}
\CF\big|_{r\to\infty}&=1,\\
\label{eq:shift_BC_infty}
\Shift^i\big|_{r\to\infty}&=({\mathbf\Omega_0}\times{\bf r})^i,\\
\label{eq:Lapse_BC_infty}
\Lapse\big|_{r\to\infty}=\CLapse\big|_{r\to\infty}&=1.
\end{align}
\end{subequations}
$\Omega_0$ is the orbital
angular velocity of a binary system, or the rotational angular
velocity of a single object, as measured at infinity.  The boundary
condition on the shift is chosen so that the time coordinate, $t^\mu =
\Lapse\Snormal^\mu + \Shift^\mu$, is helical and tracks the rotation
of the system\cite{cook96b,gourgoulhon-etal-2002a,Cook-2002}.  If we
wish to consider systems with one or more black holes, and if we
excise the interior of the black hole to avoid difficulties with
singularities, then we will also need to impose boundary conditions on
the excision surfaces.  This is the topic of the next section.

\section{Black-hole excision boundary conditions}
\label{sec:quasi-equil-bound}

The physical content of initial data depends on the choices made for
the initial-data decomposition scheme, freely specifiable data, and
the boundary conditions.  Therefore, it is important to choose
boundary conditions that are motivated by, or at least compatible
with, the sort of initial data that we wish to construct.

The first attempts to impose boundary conditions on black hole
excision boundaries were based on topological
arguments\cite{misner63,lindquist63,bowenyork80,kulkarni84}.  By
demanding that the initial-data hypersurface consist of two identical
(isometric) asymptotically flat hypersurfaces connected together at a
number of spherical excision surfaces (one for each black hole), it is
possible to show that the surfaces where the isometric sheets connect
are fixed point sets of the isometry.  This condition is enough to
determine either Dirichlet or Neumann boundary conditions at the
excision surface for any fields that are present.

Boundary conditions based on this idea have been used successfully for
generating general black-hole initial data using various initial-data
decompositions\cite{yorkpiran,choptuik-unruh-1986,cook91,cook93}.
Their first use in conjunction with the conformal thin-sandwich
decomposition\cite{gourgoulhon-etal-2002a,gourgoulhon-etal-2002b} was
only partially successful due to an unavoidable constraint violation.
The difficulties with this approach were outlined in
Ref.~\cite{Cook-2002}, where an alternative approach of using
quasiequilibrium boundary conditions was first outlined.  In this
section, we will refine, and in subsequent sections test, this
approach.

In constructing initial data on a spacelike hypersurface, we cannot
have knowledge of the event horizon that is typically used to define
the surface of a black hole.  However, we can identify the apparent
horizon of a black hole, defined as the outermost marginally
outer-trapped surface.  A marginally outer-trapped surface (MOTS), in
turn, is a surface on which the expansion, $\Oexpansion$, of the
family of outgoing null rays, $\ONull^\mu$, vanishes everywhere.

In this paper, we are interested in the situation in which each black
hole is in quasiequilibrium.  The assumptions required to enforce this
are essentially the same as those required of an ``isolated
horizon''(cf
\cite{ashtekar-etal-2000a,dreyer-etal-2003,ashtekar-krishnan-2003a}).
To ensure that the black hole is in quasiequilibrium, we enforce the
following conditions.  First, we demand that the expansion
$\Oexpansion$, vanish on the excision surface, ${\cal S}$, thus
forcing the boundary to be an apparent horizon:
\begin{equation}\label{eq:expansion_on_S}
\Oexpansion\big|_{\cal S}=0.
\end{equation}
Next, we require that the shear $\Oshear_{\mu\nu}$ of the outgoing
null rays also vanish on the excision boundary,
\begin{equation}\label{eq:shear_on_S}
\Oshear_{\mu\nu}\big|_{\cal S}=0.
\end{equation}
Consider the family of null geodesics threading the apparent
horizon to the future of our initial-data slice that are tangent to
$\ONull^\mu$ on $\cal S$. Raychaudhuri's equation for null
congruences,
\begin{equation}\label{eq:Raychaudhuri}
\LieD\ONull\Oexpansion
=-\mbox{$\frac12$}\Oexpansion^2
-\Oshear_{\mu\nu}\Oshear^{\mu\nu}+\omega_{\mu\nu}\omega^{\mu\nu}
-R_{\mu\nu}\ONull^\mu \ONull^\nu,
\end{equation}
together with Eqs.~(\ref{eq:expansion_on_S}) and (\ref{eq:shear_on_S})
are sufficient to imply that
\begin{equation}\label{eq:dexpansion_on_S}
\LieD\ONull\Oexpansion\big|_{\cal S}=0.
\end{equation}
That is, initially, the apparent horizon will evolve along
$\ONull^\mu$.  Note that $\omega_{\mu\nu}$ is the twist of the
congruence, which vanishes because the congruence is surface forming.
Also, we assume that there in no matter on ${\cal S}$, so $R_{\mu\nu}=0$.

While conditions (\ref{eq:expansion_on_S}),~(\ref{eq:shear_on_S})
and~(\ref{eq:dexpansion_on_S}) are coordinate independent, our next
and final demand breaks precisely this coordinate freedom: We demand
that the coordinate location of the apparent horizon does not move
initially in an evolution of the initial data.

As we show in the subsequent sections, the requirements listed so far
yield four conditions that can be imposed on the initial data at the
excision boundary.  However, there are {\em five} coupled initial-data
equations that must be solved in the conformal thin-sandwich approach.
When quasiequilibrium black-hole boundary conditions were first
derived, a fifth condition was considered\cite{Cook-2002}.  In
particular, the condition that $\LieD\ONull\Iexpansion=0$ was
considered, where $\Iexpansion$ is the expansion of a family of
ingoing null rays, $\INull^\mu$.  As we will show below, this fifth
condition cannot be used as a boundary condition {\em even though it
is satisfied for a stationary black hole}!

In the remainder of this section, we will derive boundary conditions
for black hole excision surfaces based on the 
demands outlined above.  A good portion of the following derivation
appeared previously\cite{Cook-2002}.  However, because of a change in
notation, and more importantly in a few sign conventions, we include
the full derivation below.

\subsection{Geometry of the excision boundary}
\label{sec:geom-excis-bound}
We demand that the excision boundary surface, ${\cal S}$, be a
spacelike 2-surface with topology $S^2$ and define $\Bnormal^i$ to be
the outward pointing unit vector normal to the surface.  In this case,
we define outward with respect to the black hole (not the domain), so
that $\Bnormal^i$ points toward infinity.  The 4-dimensional
generalization of $\Bnormal^i$ has components
$\Bnormal^\mu=[0,\Bnormal^i]$ obtained from the condition that
$\Bnormal^\mu\Snormal_\mu=0$.

The metric, $\BMetric_{ij}$, induced on ${\cal S}$ by $\SMetric_{ij}$
is given by
\begin{equation}
\BMetric_{ij} \equiv \SMetric_{ij} - \Bnormal_i\Bnormal_j.
\label{eq:bndry_metric_def}
\end{equation}
We also define the extrinsic curvature, $\BExCurv_{ij}$, of ${\cal S}$
embedded in the 3-dimensional spatial hypersurface as
\begin{equation}
\BExCurv_{ij} \equiv \BMetric^k_i\BMetric^\ell_j\SCD_{(k}\Bnormal_{\ell)}
= \mbox{$\frac12$}\BMetric^k_i\BMetric^\ell_j\LieD{\Bnormal}\BMetric_{k\ell}.
\label{eq:bndry_excurv_def}
\end{equation}

Naturally associated with ${\cal S}$ are two sets of null vectors: a
set of outgoing null rays, $\ONull^\mu$, and a set of ingoing null rays,
$\INull^\mu$, defined by
\begin{equation}
\ONull^\mu \equiv \mbox{$\frac1{\sqrt{2}}$}(\Snormal^\mu + \Bnormal^\mu) 
\quad\mbox{and}\quad
\INull^\mu \equiv \mbox{$\frac1{\sqrt{2}}$}(\Snormal^\mu - \Bnormal^\mu).
\label{eq:BNullV_def}
\end{equation}
Associated with each set of null rays is an extrinsic curvature of
${\cal S}$ as embedded in the full 4-dimensional manifold.  These are
defined as
\begin{eqnarray}
\ONExCurv_{\mu\nu} &\equiv& \mbox{$\frac12$}\BMetric^{\alpha}_{\mu}
\BMetric^{\beta}_{\nu}\LieD{\ONull}\Metric_{\alpha\beta}, 
\label{eq:ONull_ExCurv_def} \\
\INExCurv_{\mu\nu} &\equiv& \mbox{$\frac12$}\BMetric^{\alpha}_{\mu}
\BMetric^{\beta}_{\nu}\LieD{\INull}\Metric_{\alpha\beta},
\label{eq:INull_ExCurv_def}
\end{eqnarray}
where $\Metric_{\mu\nu}$ is the full spacetime metric.  Because these
tensors $\ONExCurv_{\mu\nu}$ and $\INExCurv_{\mu\nu}$ are spatial, we
will use spatial indices below.  To simplify the definitions that
follow, we will introduce various projections of $\ExCurv_{ij}$ along
and normal to the excision boundary ${\cal S}$:
\begin{eqnarray}
\PBExCurv_{ij} &\equiv& \BMetric^k_i\BMetric^\ell_j \ExCurv_{k\ell},
\label{eq:PBExCurvT_def} \\
\PBExCurv_i &\equiv& \BMetric^k_i\Bnormal^\ell \ExCurv_{k\ell},
\label{eq:PBExCurvV_def} \\
\PBExCurv &\equiv& \BMetric^{ij}\PBExCurv_{ij} = \BMetric^{ij}\ExCurv_{ij}.
\label{eq:PBExCurvS_def}
\end{eqnarray}
We can then simplify Eqs.~(\ref{eq:ONull_ExCurv_def}) and
(\ref{eq:INull_ExCurv_def}) to
\begin{equation}
\ONExCurv_{ij} = -\mbox{$\frac1{\sqrt{2}}$}(\PBExCurv_{ij} - \BExCurv_{ij})
\;\;\mbox{and}\;\;
\INExCurv_{ij} = -\mbox{$\frac1{\sqrt{2}}$}(\PBExCurv_{ij} + \BExCurv_{ij}).
\label{eq:Null_ExCurv_defs}
\end{equation}
Now, we define the expansion of outgoing null rays, $\Oexpansion$, and
ingoing null rays, $\Iexpansion$, via
\begin{eqnarray}
\Oexpansion &\equiv& \BMetric^{ij}\ONExCurv_{ij} 
= -\mbox{$\frac1{\sqrt{2}}$}(\PBExCurv - \BExCurv),
\label{eq:Oexpansion_def} \\
\Iexpansion &\equiv& \BMetric^{ij}\INExCurv_{ij} 
= -\mbox{$\frac1{\sqrt{2}}$}(\PBExCurv + \BExCurv).
\label{eq:Iexpansion_def}
\end{eqnarray}
Finally, we define the shear of the outgoing null rays, $\Oshear_{ij}$,
and ingoing null rays, $\Ishear_{ij}$, via
\begin{eqnarray} 
\Oshear_{ij} &\equiv& \ONExCurv_{ij} 
       - \mbox{$\frac12$}\BMetric_{ij}\Oexpansion,
\label{eq:Oshear_def} \\
\Ishear_{ij} &\equiv& \INExCurv_{ij} 
       - \mbox{$\frac12$}\BMetric_{ij}\Iexpansion.
\label{eq:Ishear_def}
\end{eqnarray}

\subsection{Quasiequilibrium boundary conditions}
\label{sec:excis-bound-cond}
With the definitions of Sec.~\ref{sec:geom-excis-bound}, we can now
evaluate the demands we made earlier in
Sec.~\ref{sec:quasi-equil-bound} and translate them into boundary
conditions for the conformal thin-sandwich equations.  In order to
express these as useful boundary conditions, we must write them in
terms of the variables of the conformal thin-sandwich approach.  We
must also make connection with the {\em global} notion of
quasiequilibrium that is closely associated with an approximate
helical Killing vector.

A spacetime that is in true equilibrium is said to be stationary and
has two Killing vectors of interest: a timelike Killing vector,
$\partial/\partial{t}_0$, and a spatial Killing vector associated with
rotational symmetry, $\partial/\partial\phi_0$.  If $\Omega_0$ denotes
the angular velocity of a spinning object or system as measured at infinity,
then the linear combination $\partial/\partial{t}_0 +
\Omega_0\partial/\partial\phi_0$ is referred to as the {\em helical
Killing vector}.  If a system, such as a binary, is in a state of
quasiequilibrium, there are in general no vector fields similar to
$\partial/\partial{t}_0$ or $\partial/\partial\phi_0$ that are even close
to being Killing vectors.  But there will be a helical vector field
that is an approximate Killing vector of the spacetime.  If we let
this approximate Killing vector field define our time vector,
$t^\mu$, and thus our time coordinate $t$, then we will have
$\partial/\partial{t}\approx0$ for fields in this spacetime.
Within the $\mbox{3+1}$ decomposition, we write the time vector as
\begin{equation}
t^\mu = \Lapse\Snormal^\mu + \Shift^\mu.
\label{eq:time_vec_def}
\end{equation}
Our desire for $t^\mu$ to represent an approximate helical Killing
vector is, therefore, the reason for our condition on the shift at
infinity, Eq.~(\ref{eq:shift_BC_infty}).

Now we consider the demand that the apparent horizon should initially
not move in an evolution of the quasi equilibrium initial data.
Because of Eq.~(\ref{eq:dexpansion_on_S}), the apparent horizon
initially coincides with the null surface generated by $\ONull^\mu$.
In order for the coordinates to track this null surface, the
time-vector of the evolution, $t^\mu$, must lie in this null
surface.  This requires that
\begin{equation}\label{eq:t.k=0}
t^\mu \ONull_\mu\Big|_{\cal S}=0.
\end{equation}
Substituting Eqs.~(\ref{eq:time_vec_def}) and~(\ref{eq:BNullV_def})
into Eq.~(\ref{eq:t.k=0}), and recalling that the shift vector is
spatial, $\Shift^\mu\Snormal_\mu=0$, yields
\begin{equation}\label{eq:Shift-perp-raw}
\Lapse\big|_{\cal S}=\Shift^i\Bnormal_i\big|_{\cal S}.
\end{equation}
This equation is often referred to as the Killing-horizon condition.
We split the shift vector into its component normal to the surface,
$\perpShift$, and a vector tangent to the surface, $\parShift^i$,
defined by
\begin{eqnarray}
\perpShift &\equiv& \Shift^i\Bnormal_i,
\label{eq:perpShift_def} \\
\parShift^i &\equiv& \BMetric^i_j\Shift^j.
\label{eq:parShift_def}
\end{eqnarray}
With these definitions, we see that Eq.~(\ref{eq:Shift-perp-raw}) is a
condition on the normal component of the shift, 
\begin{equation}
\perpShift\big|_{\cal S} = \Lapse\big|_{\cal S}.
\label{eq:perpShift_BC}
\end{equation}

The component of the shift {\em tangential} to the excision surface
$\cal S$, $\parShift^i$, remains unconstrained so far.  This makes
sense, because fixing the location of a surface does not restrict
motion {\em within} this surface.  We can gain insight into the
relevance of $\parShift^i$ by considering a stationary Kerr black
hole.

The Kerr spacetime has two Killing vectors of interest: A timelike
Killing vector, $\partial/\partial{t}_0$, and a spatial Killing vector
associated with rotational symmetry, $\partial/\partial\phi_0$.  The
null generators of the horizon are given by
\begin{equation}
\ONull = \frac{\partial}{\partial t_0}+\Omega_H\frac{\partial}{\partial\phi_0},
\end{equation}
where $\Omega_H$ is the angular frequency of the horizon.  If we
introduce a fiducial helical Killing vector, 
\begin{equation}
\ell=\frac{\partial}{\partial t_0}+\Omega\frac{\partial}{\partial\phi_0},
\end{equation}
for some $\Omega$, then, of course, on the horizon, 
\begin{equation}\label{eq:ell-Kerr}
\ell = \ONull + (\Omega-\Omega_H)\frac{\partial}{\partial\phi_0}.
\end{equation}
Now consider a hypersurface through Kerr to which
$\partial/\partial\phi_0$ is always tangent, e.g., the usual
Kerr--Schild slice.  To make the connection with the usual $\mbox{3+1}$
decomposition straightforward, we normalize $\ONull$ and
$\ell$ by choosing their time components to be one ($\ONull^t =
\ell^t = 1$).  If we choose the vector $\ell$ as the time vector of an
evolution, then the last term in Eq.~(\ref{eq:ell-Kerr}) corresponds
precisely to $\parShift^i$ --- this term is tangent both to the
horizon and to the hypersurface.  For the choice $\parShift^i=0$, the
last term in Eq.~(\ref{eq:ell-Kerr}) would be absent,
i.e. $\Omega=\Omega_H$.  In this case the black hole is corotating
with the coordinate system, and the generators of the horizon,
$\ONull^\mu$ do not twist relative to the helical Killing vector,
$\ell^\mu$.  Conversely, a non-rotating black hole ($\Omega_H=0$)
would have a tangential shift of
\begin{equation}
\label{eq:parShift-nonrotating}
\parShift^i=\Omega\left(\frac{\partial}{\partial\phi_0}\right)^i.
\end{equation}

For a binary black hole in quasiequilibrium, neither $\partial/\partial
t_0$ nor $\partial/\partial\phi_0$ exist as separate Killing vectors.
However, based on the discussion above, we expect that $\parShift^i$
is still connected to the rotation of the black hole.  In the case
$\parShift^i=0$, the horizon generators have no twist relative to the
helical Killing vector (which coincides with the time-vector),
corresponding to corotating black holes.  Any rotation,
$\parShift^i\propto \left(\partial/\partial\phi_{\cal S}\right)^i$, where
$\left(\partial/\partial\phi_{\cal S}\right)^i$ lies in the surface
${\cal S}$ would impart additional rotation on the black hole. Below we
will make these notions precise. 

Having obtained a boundary condition on $\perpShift$, we now turn our
attention to 
Eqs.~(\ref{eq:expansion_on_S}) and~(\ref{eq:shear_on_S}).  We need to
consider the horizon boundary ${\cal S}$ in the conformal space.  The
conformal transformation on $\SMetric_{ij}$ (\ref{eq:Metric_decomp})
induces a natural conformal weighting for $\BMetric_{ij}$ and for the
unit normal to ${\cal S}$,
\begin{eqnarray}
\BMetric_{ij} &\equiv& \CF^4\CBMetric_{ij},
\label{eq:CBMetric_def} \\
\Bnormal^i &\equiv& \CF^{-2}\CBnormal^i.
\label{eq:CBnormal_def}
\end{eqnarray}
If we also define $\CBCD^i$ as the covariant derivative compatible
with $\CBMetric_{ij}$, then without loss of generality, we can
express the expansion of the outgoing null rays, $\Oexpansion$,
as
\begin{equation}
\Oexpansion = {\CF^{-2}\over\sqrt2}\left(
          \CBMetric^{ij}\CCD_i\CBnormal_j
	  + 4 \CBnormal^k\CCD_k\ln\CF
	  - \CF^2\PBExCurv\right),
\label{eq:Oexp_gen}
\end{equation}
and the shear of the outgoing null rays, $\Oshear^{ij}$, as
\begin{eqnarray}
\Oshear^{ij} &=& \frac1{\sqrt{2}}
(\BExCurv^{ij} - \mbox{$\frac12$}\BMetric^{ij}\BExCurv)
\left(1 - \frac{\perpShift}{\Lapse}\right) 
\nonumber \\ && \mbox{}
- \frac1{\sqrt{2}}\frac{\CF^{-4}}{\Lapse}\left[
\CBCD^{(i}\parShift^{j)} - \mbox{$\frac12$}\CBMetric^{ij}\CBCD_k\parShift^k
\right. \label{eq:Oshear_gen} \\
&& \left.\hspace{0.75in}\mbox{}
- \mbox{$\frac12$}(\CBMetric^i_k\CBMetric^j_\ell\CMtd^{k\ell}
- \mbox{$\frac12$}\CBMetric^{ij}\CBMetric_{k\ell}\CMtd^{k\ell})\right].
\nonumber
\end{eqnarray}

It is now clear how to obtain the remaining boundary conditions.  By
applying condition~(\ref{eq:expansion_on_S}) to Eq.~(\ref{eq:Oexp_gen}),
we obtain a boundary condition that forces an excision boundary to
be an apparent horizon (or MOTS).  The condition is
\begin{equation}
\CBnormal^k\CCD_k\ln\CF\Big|_{\cal S} = 
       -\frac14\left(\CBMetric^{ij}\CCD_i\CBnormal_j 
                - \CF^2\PBExCurv\right)\Big|_{\cal S},
\label{eq:AH_BC}
\end{equation}
and it takes the form of a nonlinear Robin-type boundary condition on
the conformal factor, $\CF$.  Finally, if we recall the condition from
Eq.~(\ref{eq:perpShift_BC}) and that we have chosen
$\CMtd^{ij}=0$ everywhere, and apply
condition~(\ref{eq:shear_on_S}) to Eq.~(\ref{eq:Oshear_gen}), we
obtain a condition that restricts the form of $\parShift^i$.
The condition is that
\begin{equation}
\CBCD^{(i}\parShift^{j)}\Big|_{\cal S}
- \mbox{$\frac12$}\CBMetric^{ij}\CBCD_k\parShift^k\Big|_{\cal S} = 0.
\label{eq:parShift_BC}
\end{equation}
This shows that the components of the shift that are associated with
the spin of the black hole must be proportional to a conformal Killing
vector of the conformal metric, $\CBMetric_{ij}$, defined on the
2-dimensional excision surface.  

This condition is quite remarkable.  
Recall that any 2-surface that is topologically $S^2$ is conformally
equivalent to the unit 2-sphere.  If you consider a unit 2-sphere
embedded in a flat 3-dimensional Euclidean space, then there is a
family of rotational Killing vectors, $\xi^i$, associated with any
rotation axis passing through the center of the 2-sphere.  Because
$\xi^i\hat{n}_i=0$ on the 2-sphere, where $\hat{n}^i$ is the unit
normal vector on the 2-sphere, we see that $\xi^i$ trivially form a
family of 2-dimensional vectors tangent to the 2-sphere and that these
are Killing vectors of the metric on the unit 2-sphere.  But, the
Killing vectors associated with any metric are also conformal Killing
vectors of any metric conformally related to it.  So, if $\varphi$
represents a conformal transformation such that
$\varphi^4\CBMetric_{ij}$ is the metric of the unit 2-sphere, then
$\xi^i$ will satisfy the conformal Killing equation on
$\CBMetric_{ij}$.  Thus,
\begin{equation}
\parShift^i = \Omega_r \xi^i
\label{eq:Killing_shift_def}
\end{equation}
will satisfy Eq.~(\ref{eq:parShift_BC}), with $\Omega_r$ being an
arbitrary parameter.  The freedom left in
Eq.~(\ref{eq:Killing_shift_def}) is precisely what is required to
parameterize an arbitrary
spin on the black hole.  The parameter $\Omega_r$ is associated with
the magnitude of the rotation or spin of the black hole, whereas the
axis of rotation of $\xi^i$ is related to the orientation of the spin.
Of course, $\Omega_r$ does not correspond directly to the rotational
angular velocity of the black hole.  From the discussion leading to
Eq.~(\ref{eq:parShift-nonrotating}), it is clear that $\Omega_r=0$
corresponds to a black hole that is corotating with an approximate
helical Killing vector $t^\mu$ and thus represents a black hole that
is rotating as seen by an inertial observer at infinity.  In order to
construct a black hole that is not rotating as seen from infinity, we
need to choose a shift that is similar in form to
Eq.~(\ref{eq:parShift-nonrotating}).  It seems reasonable to choose
$\Omega_r = \Omega_0$ and to pick the conformal Killing vector, $\xi^i$
from the family that corresponds to rotation about an axis that is
perpendicular to the plane of the orbit.

To summarize, the quasiequilibrium conditions defined in
Eqs.~(\ref{eq:expansion_on_S}) and~(\ref{eq:shear_on_S}) define
boundary conditions on the conformal factor, $\CF$, via
Eq.~(\ref{eq:AH_BC}) and on the shift vector, $\Shift^i$, via
Eqs.~(\ref{eq:perpShift_def}), (\ref{eq:parShift_def}),
(\ref{eq:perpShift_BC}), and (\ref{eq:Killing_shift_def}).  These
total to four of the five necessary boundary conditions for solving
the coupled elliptic equations associated with the conformal
thin-sandwich equations.  Missing is a condition on the conformal
lapse, $\CLapse$.

\subsection{Boundary conditions on the lapse function}
\label{sec:bound-cond-lapse}
A possible boundary condition on the lapse was derived in
Ref.~\cite{Cook-2002}.  The condition was essentially based on the
reasonable quasiequilibrium condition that
$\LieD{\ONull}\Iexpansion=0$.  Notice that this is the change in the
expansion of {\em ingoing} null rays, $\INull^\mu$, with respect to 
the {\em outgoing} null congruence.  The
resulting boundary condition takes the form
\begin{equation}
\alpha_{BC}\equiv\left(\PBExCurv\CBnormal^i\CCD_i\Lapse -
           {\cal E}\Lapse\right)\Big|_{\cal S}=0,
\label{eq:QE_lapse_BC}
\end{equation}
where ${\cal E}$ is a nonlinear operator that is elliptic within the
surface ${\cal S}$ (see Eq.~(84) of Ref.~\cite{Cook-2002} for a
precise description).  This condition is satisfied on the horizon
of a stationary black hole and it seemed to supply a reasonable
boundary condition for the lapse to be used in conjunction with
the previously defined boundary conditions for the conformal factor
and shift vector.

We implemented the full set of boundary conditions within the code
described in Ref.~\cite{Pfeiffer-etal-2002a} and attempted to solve
the full set of conformal thin-sandwich equations for the case of a
single nonrotating black hole.  It became clear immediately that the
iterative solutions would not converge in general.  Interestingly, if
the analytic solution for an isolated black hole was supplied for the
starting point of the iterations, then the equations and boundary
conditions were satisfied to truncation error, but the iterations were
at best only marginally stable.  Furthermore, if any one of the
boundary conditions on $\CF$, $\perpShift$, or $\Lapse$ were replaced
by an arbitrary Dirichlet or Neumann boundary condition, then the
iterative solution was convergent to a solution representing a static
black hole {\em and the omitted boundary condition was satisfied}.

This clearly indicates that the set of boundary conditions
including Eq.~(\ref{eq:QE_lapse_BC}) is degenerate and leads to an
ill-posed elliptic system.  We can understand the nature of the
degeneracy by considering the family of time-independent maximal
slicings of
Schwarzschild\cite{reinhart-1973,estabrook-etal-73,beig-omurchadha-1998}.
The line element for the spatial metric, lapse, and shift vector are
\begin{subequations}
\begin{eqnarray}
\td{s}^2 &=& \left(1 - \frac{2M}{R} + \frac{C^2}{R^4}\right)^{-1}\td{R}^2
              + R^2\td{\Omega}^2,
\label{eq:Schwz_max_metric} \\
\Lapse &=& \sqrt{1 - \frac{2M}{R} + \frac{C^2}{R^4}},
\label{eq:Schwz_max_lapse} \\
\Shift^R &=& \frac{C}{R^2}\sqrt{1 - \frac{2M}{R} + \frac{C^2}{R^4}},
\label{eq:Schwz_max_shift}
\end{eqnarray}
where $R$ is the usual Schwarzschild ``areal'' radial coordinate, $M$
is the mass of the black hole, and $C$ is a constant parametrizing the
family of maximal slicings.  In spherical coordinates, the extrinsic
curvature takes the form
\begin{equation}
\ExCurv^i_j = \frac{C}{R^3}\left[
  \begin{array}{ccc}
    -2 &  0 &  0 \\
    0 & 1 &  0 \\
    0 &  0 & 1
    \end{array}
  \right].
\label{eq:Schwz_max_excurv}
\end{equation}
\end{subequations}
For $0\le C/M^2 <\sqrt{\frac{27}{16}}\approx 1.299$ the maximal spatial slice
extends from spatial infinity, through the black-hole interior, and to
the second spatial infinity of the maximally extended Schwarzschild
geometry.  When $C/M^2=0$, we recover the standard Schwarzschild maximal
slice that passes through the bifurcation point.  And, for
$C/M^2 > \sqrt{\frac{27}{16}}$, the maximal spatial slice extends from
spatial infinity, through the black-hole horizon, and ends on the
singularity.

If desired, the family of maximal slicings of Schwarzschild can be be
rewritten in terms of an isotropic radial coordinate, $r$.  It is then easy
to verify that the boundary conditions (\ref{eq:AH_BC}),
(\ref{eq:perpShift_BC}), and (\ref{eq:QE_lapse_BC}) are satisfied on 
the horizon for any value of $C/M^2$.  So, we see that while the 
set of boundary conditions proposed in Ref.~\cite{Cook-2002} hold for
a time-independent configuration, they do not uniquely fix the spatial
slicing.  It is in this way that they are degenerate.

If we consider the value of the lapse on the horizon, we find that
\begin{equation}
\Lapse(r_{\cal H}) = \frac14\left(\frac{C}{M^2}\right),
\label{eq:Schwz_Dir_lapse}
\end{equation}
Where $r_{\cal H}$ denotes the location of the horizon in isotropic
coordinates.  Note that $r_{\cal H}\ne M/2$ unless $C/M^2=0$.  If,
instead of using the lapse boundary condition (\ref{eq:QE_lapse_BC}),
we simply fix a Dirichlet value for the lapse, then we find that we
have effectively chosen a value of $C/M^2$ and thus a particular
maximal slicing of Schwarzschild.  Similarly, we find that the family
of mixed boundary conditions
\begin{equation}
\left.{\partial\Lapse\over\partial r}\right|_{r_{\cal H}} =
       {\cal A} \left.{\Lapse\over r}\right|_{r_{\cal H}}
\label{eq:Schw_mixed_lapse}
\end{equation}
corresponds to a slicing choice of $C/M^2 = \sqrt{{\cal A}^2 +
4}-{\cal A}$, where ${\cal A}$ is any real number.  It is clear that
any reasonable choice of a Dirichlet, Neumann, or mixed boundary
condition on the lapse will uniquely fix a particular maximal slicing
of Schwarzschild and effectively break the degeneracy.

As mentioned previously, our numerical investigations have shown that
we could also have chosen to use the lapse boundary condition
(\ref{eq:QE_lapse_BC}) and instead fix either $\CF$ or $\perpShift$ on
the horizon via a Dirichlet, Neumann, or mixed boundary condition.
One reason to choose to replace the lapse condition
(\ref{eq:QE_lapse_BC}), as opposed to the boundary conditions on the
conformal factor (\ref{eq:AH_BC}) or the shift
(\ref{eq:perpShift_BC}), is that the lapse boundary condition
is much more complex.  However, there is a more fundamental reason to
choose to replace the lapse condition.  The degeneracy that we must
eliminate is in the choice of the spatial slice which is a choice of
the initial temporal gauge.  The lapse function fixes the evolution of
the temporal gauge.  Therefore it seems reasonable that we should
consider the choice of the lapse boundary condition as part of the
initial temporal gauge choice.  It is customary to view the choice
of the trace of the extrinsic curvature as fixing the initial temporal
gauge.  However, this is apparently not sufficient within the
conformal thin-sandwich approach when interior boundaries are present.

This last assertion, that the lapse boundary condition must be chosen
as part of the initial temporal gauge choice, is supported by the
behavior of a very special solution of Einstein's equations -- the
static Schwarzschild solution.  However the same generic behavior
is seen in a broad class of examples as we will outline below.

\section{Quasiequilibrium solutions for a single black hole}
\label{sec:quas-solut-single}

As we have seen in Sec.~\ref{sec:conf-thin-sandw}, the conformal thin
sandwich equations require specification of free data, which are the
conformal metric $\CMetric_{ij}$ and its time derivative $\CMtd_{ij}$,
as well as the mean curvature $\TrExCurv$ and its time derivative
$\partial_t\TrExCurv$.  Moreover, boundary conditions are required on
the variables being solved for, the conformal factor $\CF$, the shift
$\Shift^i$ and the lapse $\Lapse$.  The quasiequilibrium
approximation fixes a good portion of these choices, namely
$\CMtd_{ij}=\partial_t\TrExCurv=0$, as well as the the following
boundary conditions at the excised regions: The apparent horizon
condition Eq.~(\ref{eq:AH_BC}) on $\CF$, the null horizon condition
Eq.~(\ref{eq:perpShift_BC}) on $\perpShift$, and
Eq.~(\ref{eq:Killing_shift_def}) on $\parShift^i$.  At spatial
infinity, the boundary conditions are straightforward and are given by
Eqs.~(\ref{eq:outer-BCs}).

Before the conformal thin-sandwich equations can be solved, we have to
choose the remaining quantities which are not fixed by the
quasiequilibrium framework.  These are $\CMetric_{ij}$, $\TrExCurv$,
and an inner boundary condition on the lapse $\Lapse$.  Furthermore,
we have to choose the shape of the excised regions, $\cal S$.  As
argued above, and as we confirm below, the lapse boundary condition
and $\TrExCurv$ are part of the temporal gauge choice. It thus remains
to choose $\CMetric_{ij}$ and the shape of the excised regions, $\cal
S$.

In this work, we will assume that the conformal three-geometry is
flat, and we will always excise exact spheres.  These choices are not
motivated by physical considerations, and they will affect the quality
of the quasiequilibrium solutions we obtain in this paper.  For
example, the Kerr spacetime does not admit conformally flat slices
\cite{garat-price-2000, Kroon:2004}.  Therefore, when we solve for
a rotating black hole, our initial-data sets will not exactly represent
a Kerr black hole, but will rather correspond to a perturbed Kerr
black hole, which will settle down to Kerr.  We stress that this
failure of the initial-data sets constructed here to represent Kerr is
not inherent in the quasiequilibrium method, but is caused by our
choices for $\CMetric_{ij}$ and $\cal S$.  With the appropriate
choices for $\CMetric_{ij}$ and $\cal S$, the quasiequilibrium method
can reproduce exactly any time-independent solution of Einsteins
equations.  Indeed, for {\em single} black holes, better choices for
$\CMetric_{ij}$ and $\cal S$ are easily obtained from stationary
analytic solutions of Einstein's equations, for example based on
Kerr--Schild coordinates.  While such a choice certainly leads to
single black hole initial-data sets closer to the true Kerr metric, it
is not clear how to generalize to {\em binary} black hole
configurations.  A widely used approach superposes single black hole
quantities to construct binary black hole initial data (e.g.,
\cite{cook91,cook94a,Kerrbinary98,Kerrbinary2000,Yo-etal:2004}).
However, because of the nonlinear nature of Einsteins equations, and
since often, the black holes are separated by only a few
Schwarzschild radii, the superposition introduces uncertainties that
may be large \cite{Pfeiffer-etal-2002a} and that have not yet been
adequately quantified.  In this paper, rather than using
superposition, we start with choices for $\CMetric_{ij}$ and $\cal S$
that are not optimal for single black hole spacetimes, but that are
equally well suited for binary black hole configurations.  We then use
the single black hole solutions to quantify the effects of our
approximation.

We solve the conformal thin-sandwich equations with the
pseudo-spectral collocation method described in \cite{Pfeiffer2003}.
For the single black hole spacetimes, typically, two spherical shells
are employed.  The inner one ranges from the excised sphere to a
radius of $\sim 20$ and distributes grid-points exponentially in
radius.  The outer shell has an outer radius of typically $\sim
10^{10}$, and employs an inverse mapping in radius, which is well
adapted to the $1/r$--falloff of many quantities.  The fifth elliptic
equation (for the lapse function) is coded in the form of
Eq.~(\ref{eq:Const_TrK_eqn}), i.e. as an equation for
$\CF^7\CLapse=\Lapse\CF$.  Therefore, we formulate the lapse boundary
condition as a condition on $\Lapse\CF$.  Finally, we note that we
always solve the three-dimensional initial value equations, even in
cases which have spherical or cylindrical symmetries like the single
black hole solutions.

\subsection{Spherical symmetry} 
\label{Sec:SingleBH_Sphere}

We begin by solving for spherically symmetric initial-data sets that
contain one black hole.  In spherical symmetry, the assumption of
conformal flatness is no restriction, because any spherically
symmetric metric can be made conformally flat through an appropriate
radial coordinate transformation.  For example, a hypersurface through the
Schwarzschild spacetime of constant Kerr--Schild coordinate time 
has the induced metric 
\begin{equation}\label{eq:KS-spatial}
ds^2 = \left(1+\frac{2M}{R}\right)dR^2+R^2d\Omega^2,
\end{equation}
where $R$ denotes the areal radius.  Here, the
coordinate transformation \cite{Cook-2002}
\begin{equation}\label{eq:KS-flat-r}
r=\frac{R}{4}\left(1+\sqrt{1+\frac{2M}{R}}\right)^2
e^{2-2\sqrt{1+2M/R}},
\end{equation}
brings the induced metric into conformally flat form,
\begin{equation}
ds^2=\CF_{\mbox{\tiny KS}}^4\left(r^2dr^2+r^2d\Omega^2\right),
\end{equation}
with $\CF_{\mbox{\tiny KS}}=\sqrt{R/r}$.

In order to support our claim that $\TrExCurv$ and the lapse boundary
condition merely represent a coordinate choice, we solve the
quasiequilibrium equations for several, essentially arbitrary,
choices for these quantities.  For the mean curvature, we choose
\begin{subequations}
\begin{align}
\label{eq:K-1}
\TrExCurv&=0,\\
\TrExCurv&=\frac{2M}{r^2},\\
\label{eq:K-3}
\TrExCurv&=K_{\mbox{\tiny KS}}\equiv\frac{2M}{R^2}\left(1+\frac{2M}{R}\right)^{-3/2}
           \left(1+\frac{3M}{R}\right),
\end{align}
\end{subequations}
where in the last case, $R$ is given implicitly by
Eq.~(\ref{eq:KS-flat-r}).  Equation~(\ref{eq:K-3}) represents
the mean curvature for a Kerr--Schild slice of the Schwarzschild
spacetime with mass $M$.

For the lapse boundary condition at the excised spheres, we use
\begin{subequations}
\begin{align}
\label{eq:Lapse-BC-1}
\frac{d\Lapse\CF}{dr}\bigg|_{\cal S}&=0,\\
\label{eq:Lapse-BC-2}
\frac{d\Lapse\CF}{dr}\bigg|_{\cal S}&=\frac{\Lapse\CF}{2r}\bigg|_{\cal S},\\
\label{eq:Lapse-BC-3}
\Lapse\CF\big|_{\cal S}&=\frac{1}{2},\\
\label{eq:Lapse-BC-4}
\Lapse\CF\big|_{\cal S}&=\frac{1}{\sqrt{2}}\CF_{\mbox{\tiny KS}}\big|_{\cal S}.
\end{align}
\end{subequations}
The last condition, Eq.~(\ref{eq:Lapse-BC-4}) is correct for the
Kerr--Schild slice. 

We now compute twelve initial-data sets, combining any of the choices
for $\TrExCurv$ with any of the lapse boundary conditions.  In order
to fully recover the Kerr--Schild slice for the choices (\ref{eq:K-3})
and~(\ref{eq:Lapse-BC-4}), we set the radius of the excised sphere
to 
\begin{equation}
r_{exc}=r\big|_{R=2M}=\frac{1}{2}\left(1+\sqrt{2}\right)^2e^{2-2\sqrt{2}}M.
\end{equation}

For each of the initial-data sets, we compute the residual in the
Hamiltonian and momentum constraints,
Eqs.~(\ref{eq:Hamiltonian_const}) and~(\ref{eq:Momentum_const}).  We
compute ADM quantities of the initial-data set by the standard
integrals at infinity in Cartesian coordinates,
\begin{align}\label{eq:EADM}
E_{\mbox{\tiny ADM}}&=\frac{1}{16\pi}\int_{\infty} 
  \left(\gamma_{ij,j}-\gamma_{jj,i}\right)\,d^2S_i,\\
\label{eq:JADM}
J_{(\xi)}&=\frac{1}{8\pi}\int_{\infty}
  \left(K^{ij}-\gamma^{ij}K\right)\xi_j\,d^2S_i.
\end{align}
For the $x$-component of the linear ADM-momentum, $\xi=\hat e_x$ in
Eq.~(\ref{eq:JADM}).  The choice $\xi=x\hat e_y-y\hat e_x$ yields the
$z$-component of the ADM-like angular momentum as defined by York
\cite{york79}.  We also compute the irreducible mass 
\begin{equation}
M_{\rm irr}=\sqrt{\frac{A_{AH}}{16\pi}},
\end{equation}
where we have approximated the area of the (unknown) event horizon by
the area $A_{AH}$ of the apparent horizon, and the Komar mass,
\begin{equation}
M_{\mbox{\tiny K}}=\frac{1}{4\pi}\int_{r=\infty}\SMetric^{ij}\left(\SCD_i\Lapse-\Shift^k\ExCurv_{ik}\right)d^2S_j.
\end{equation}
Finally, we evaluate the time-derivative of $\ExCurv_{ij}$ by
Eq.~(\ref{eq:ExCurv_evol}), evaluate the residual $\Lapse_{BC}$ of the
quasiequilibrium lapse condition, Eq.~(\ref{eq:QE_lapse_BC}), and
compute
\begin{equation}
\partial_t\ln\CF=-\frac{1}{6}\left(\Lapse\TrExCurv-\SCD_i\Shift^i\right),
\end{equation}
which follows from the trace of Eq.~(\ref{eq:metric_evol}).

\begin{figure}
\centerline{\includegraphics[scale=0.465]{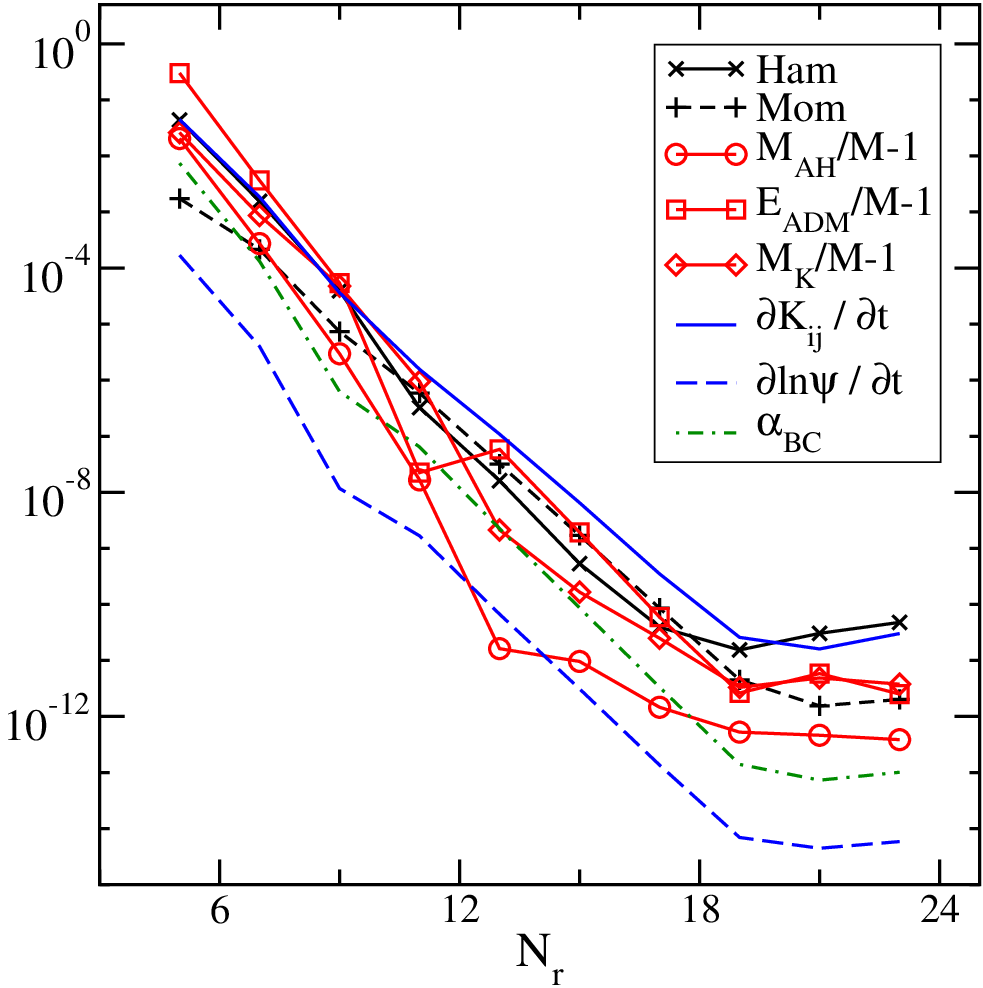}}
\caption{\label{fig:SolveKS}Solution of the quasiequilibrium
equations recovering the Kerr--Schild slicing of Schwarzschild.
Plotted are the maximum values of Hamiltonian and momentum constraints
and of time-derivatives, as well as the deviation of $M_{\rm irr}$,
$E_{\mbox{\tiny ADM}}$ and $M_{K}$ from the analytical answer $M$.  $N_r$ is the
radial number of collocation points in each of the two spherical
shells.}
\end{figure}  

Figure~\ref{fig:SolveKS} presents a convergence plot for one of the
twelve cases, Eqs.~(\ref{eq:K-3}) and~(\ref{eq:Lapse-BC-4}).  This
case recovers the usual Kerr--Schild slice with mass $M$.  The residual
of the Hamiltonian and momentum constraints decrease exponentially
with resolution, and the three different masses $M_{\rm irr}$, $E_{\mbox{\tiny ADM}}$
and $M_{\mbox{\tiny K}}$ all converge to the expected result, $M$.  Furthermore, the
time-derivatives exponentially converge to zero, and ADM linear and
angular momenta converge to zero, too.  The vanishing time-derivatives
indicate that the quasiequilibrium method constructs lapse and shift
along the timelike Killing vector of the Schwarzschild spacetime.

For different choices of $\TrExCurv$ or for the lapse BC, we find that
the masses are no longer exactly unity.  However, for {\em all}
choices of $\TrExCurv$ and lapse BC, we find to within truncation
error, that the three masses agree,
\begin{equation}
M_{\rm irr}=E_{\mbox{\tiny ADM}}=M_{\mbox{\tiny K}},
\end{equation}
and that all time-derivatives (and the lapse condition
Eq.~[\ref{eq:QE_lapse_BC}]) vanish:
\begin{equation}
\partial_t\psi=\partial_tK_{ij}=\alpha_{BC}=0.
\end{equation}
These findings are summarized in Table~\ref{tab:singleBHs}.  These
runs indicate that any (reasonable) choice for the mean curvature
$\TrExCurv$ and the lapse BC recovers a slice through
Schwarzschild with time-vector along the timelike Killing vector.

\begin{table}
\caption{\label{tab:singleBHs}Spherically symmetric quasiequilibrium
initial-data sets.  Given are the irreducible mass, ADM-energy and
Komar mass (these three quantities are found to be identical to within
truncation error).  The last column gives an upper bound on the
deviation from zero of Hamiltonian and momentum constraints,
time-derivatives of $\CF$ and $\ExCurv_{ij}$, as well as the lapse
condition $\Lapse_{BC}$}
\begin{ruledtabular}
\begin{tabular}{cl@{\hspace*{-3.5em}}dc}
%
%
$\TrExCurv$ & Lapse BC & 
\multicolumn{1}{c}{\hspace*{0.9cm}$
M_{\rm irr},E_{\mbox{\tiny ADM}},M_{\mbox{\tiny K}}
$} 
 & ${\cal H}, {\cal M}_i$,  ``$\partial_t$'' \\
$0$ & $(\Lapse\CF)'=0$              & 1.48079275 & $<10^{-10}$     \\
    & $(\Lapse\CF)'=\Lapse\CF/(2r)$ & 1.61967937 & $<10^{-10}$     \\
    & $\Lapse\CF=1/2$               & 1.65726413 & $<10^{-10}$     \\
    & $\Lapse\CF=(\Lapse\CF)_{\mbox{\tiny KS}}$  & 1.28974831 & $<10^{-10}$     \\[.75em]
$2M/r^2$  & $(\Lapse\CF)'=0$      & 0.68281    & $<10^{-9}$      \\
    & $(\Lapse\CF)'=\Lapse\CF/(2r)$ & 0.73571    & $<10^{-9}$      \\
    & $\Lapse\CF=1/2$               & 0.99176    & $<10^{-8}$      \\
    & $\Lapse\CF=(\Lapse\CF)_{\mbox{\tiny KS}}$  & 0.77233    & $<10^{-9}$      \\[.75em]
$K_{\mbox{\tiny KS}}$ & $(\Lapse\CF)'=0$         & 0.9942475  & $<10^{-9}$      \\
    & $(\Lapse\CF)'=\Lapse\CF/(2r)$ & 1.091637   & $<10^{-9}$      \\
    & $\Lapse\CF=1/2$               & 1.295099   & $<10^{-9}$      \\
    & $\Lapse\CF=(\Lapse\CF)_{\mbox{\tiny KS}}$  & 1.0000000  & $<10^{-10}$      \\
\end{tabular}
\end{ruledtabular}
\end{table}

For the maximal slices, $\TrExCurv=0$, the different lapse BCs choose
different parameters $C/M^2$ in the family of maximal slicings,
Eqs.~(\ref{eq:Schwz_max_metric})--(\ref{eq:Schwz_max_excurv}).  The
boundary conditions in Eqs.~(\ref{eq:Lapse-BC-1})--(\ref{eq:Lapse-BC-4})
correspond, respectively, to $C/M^2\!=\!2/3(\sqrt{13}-1)$,
$C/M^2\!=\!4/3$, $C/M^2\approx 1.2393$, and $C/M^2\approx 2.4905$.
Based on the results of Table~\ref{tab:singleBHs}, we conjecture that
for any (reasonable) function $\TrExCurv(r)$, there exists a
one-parameter family of spherically symmetric slicings, which extend
from the horizon to spatial infinity.

In situations with less symmetry like binary black holes, we prefer
Neumann or Robin boundary conditions on the lapse
(Eqs.~[\ref{eq:Lapse-BC-1}] and~[\ref{eq:Lapse-BC-2}]), because they
allow the lapse on the horizon to respond to tidal deformations.
Furthermore, as Table~\ref{tab:singleBHs} confirms that the choice of
mean curvature plays a marginal role, we will concentrate on the most
obvious choice, maximal slicing $\TrExCurv=0$, below.

In summary, the quasiequilibrium method is singularly successful for
spherically symmetric spacetimes: There exists a natural choice for
$\CMetric_{ij}$ (the flat metric), that, together with {\em any}
(reasonable) choices for $\TrExCurv$ and the lapse-boundary condition
yields a slice though Schwarzschild {\em and} the timelike Killing
vector that results in a completely time-independent evolution.

\subsection{Single rotating black holes}

In Sec.~\ref{sec:quasi-equil-bound} we explained that the tangential
component of the shift-vector on the excised surface induces a
rotation on the hole, cf. Eq.~(\ref{eq:Killing_shift_def}).  We now
test this assertion by constructing initial-data sets for single
rotating black holes.

\begin{figure}
\centerline{\includegraphics[scale=0.465]{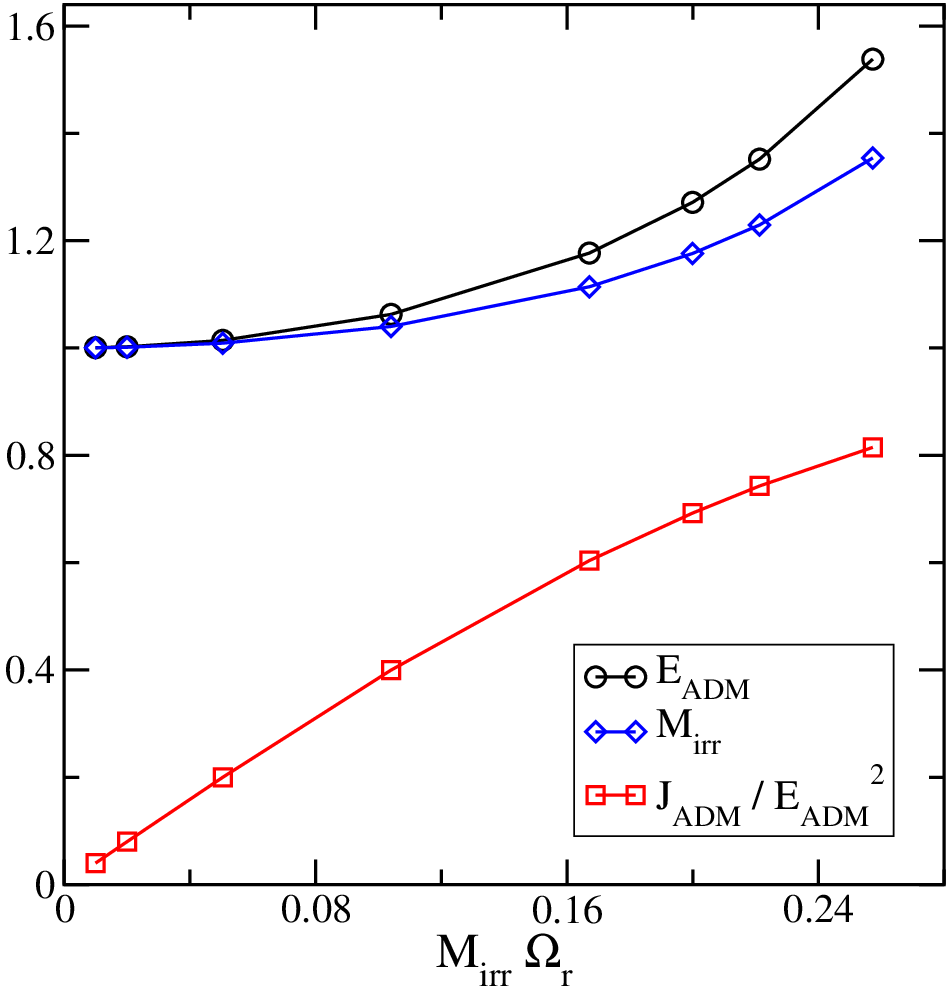}}
\caption{\label{fig:Spinning2} Spinning single black hole initial-data
sets.  }
\end{figure}

We set the mean curvature and the lapse BC by Eqs.~(\ref{eq:K-1})
and~(\ref{eq:Lapse-BC-1}), and we continue to use the conformal
flatness approximation, $\CMetric_{ij}=\FMetric_{ij}$, and choose the
excised region to be a coordinate sphere centered on the origin with
radius $r_{exc}=0.8594997$.  This value ensures a unit-mass black
hole in the limit of no rotation for the choice of $\TrExCurv$ and
lapse BC.  The shift boundary conditions encode the rotation.  At
infinity, we set $\beta^i=0$; on the horizon,
Eq.~(\ref{eq:Killing_shift_def}) implies
\begin{equation}\label{eq:shift-BC-singlerotating}
\parShift^i = \Omega_r^j x^k \epsilon^{ijk},
\end{equation}
where $x^k$ is the Cartesian coordinate separation of points on $\cal
S$ to the center of the excised sphere.  We choose $\Omega^i_r$
parallel to the z-axis, and solve the initial value equations for
different magnitudes of $\Omega_r$.  For each solution, we compute the
diagnostics mentioned in Sec.~\ref{Sec:SingleBH_Sphere}.
Figure~\ref{fig:Spinning2} presents the ADM-energy, irreducible mass and
angular momentum of the obtained data sets.  We see that, for small
$\Omega_r$, the angular momentum increases linearly with $\Omega_r$, 
as expected.

As discussed early in Sec.~\ref{sec:quas-solut-single}, our assumption
of conformal flatness will necessarily introduce some errors when
solving for a rotating black hole, because the Kerr metric does not
admit conformally flat slices\cite{garat-price-2000, Kroon:2004}.
Very interesting are therefore measures of the deviation of the
quasiequilibrium initial-data sets to a slice through the exact Kerr
spacetime.

\begin{figure}
\centerline{\includegraphics[scale=0.465]{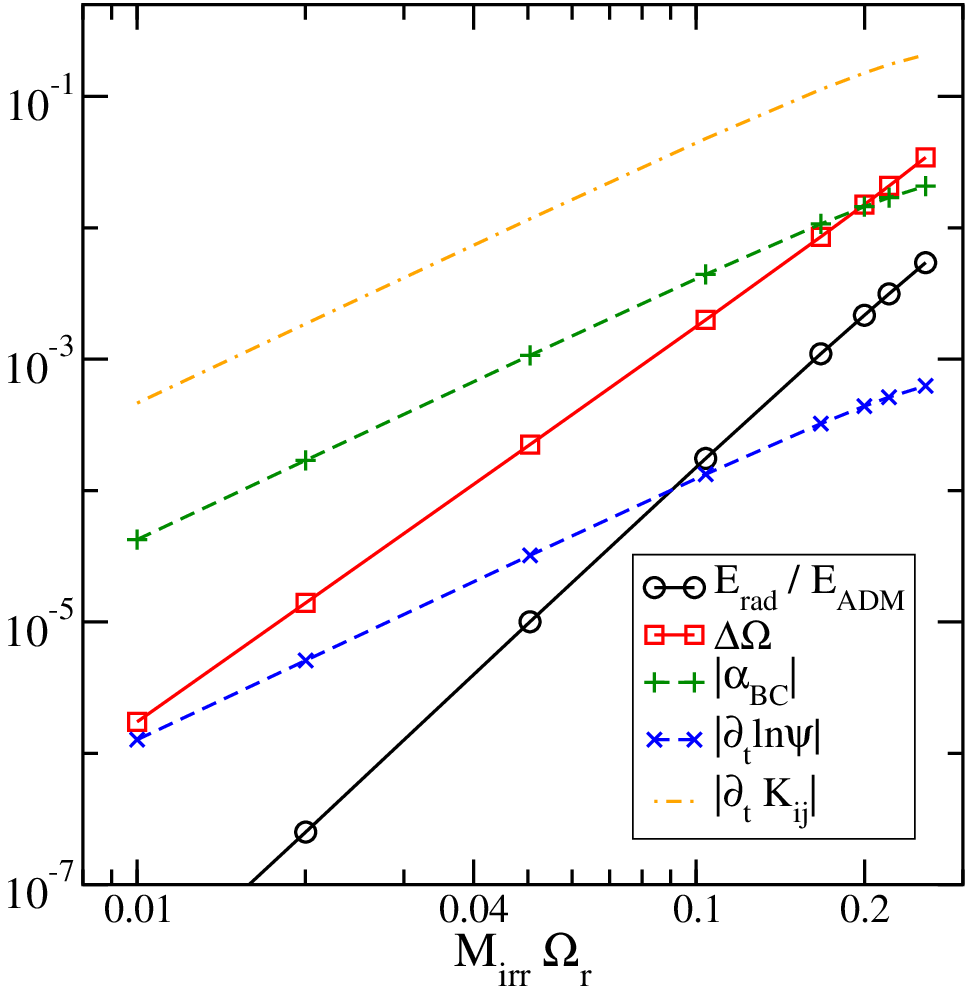}}
\caption{\label{fig:Spinning1}Single rotating black hole initial-data
sets: Deviations from the exact Kerr metric.}
\end{figure}

One such quantity is the maximum amount of energy that can potentially
be radiated to infinity,
\begin{equation}
E_{rad} = \sqrt{E_{\mbox{\tiny ADM}}^2 - P_{\mbox{\tiny ADM}}^2} - \sqrt{M_{\rm irr}^2 +
\frac{J_{\mbox{\tiny ADM}}^2}{4 M_{\rm irr}^2}}.
\end{equation}
For a stationary spacetime, $E_{rad}=0$.  Another interesting question
is how closely $\Omega_r$ of Eq.~(\ref{eq:shift-BC-singlerotating})
corresponds to the angular frequency of the horizon.  For a Kerr black
hole with angular momentum $J_{\mbox{\tiny ADM}}$ and total mass $E_{\mbox{\tiny ADM}}$, the
angular frequency of the horizon is given by \cite{MTW}
\begin{equation}
\Omega_{H} = \frac{J_{\mbox{\tiny ADM}}/E^3_{\mbox{\tiny ADM}}}{2+2\sqrt{1-\left(J_{\mbox{\tiny ADM}}/E_{\mbox{\tiny ADM}}^2\right)^2}},
\end{equation}
so that
\begin{equation}
\Delta\Omega \equiv E_{\mbox{\tiny ADM}}\left(\Omega_{r} - \Omega_{H}\right)
\end{equation}
measures the deviation of $\Omega_r$ from the angular frequency of the
horizon.  Figure~\ref{fig:Spinning1} presents these quantities.  The
maximum radiation content, $E_{rad}$, is proportional to $(M_{\rm
irr}\Omega_r)^4$.  For the binary black hole data sets we construct
below in Sec.~\ref{sec:quasi-circ-orbits}, the relevant angular
frequency is the orbital angular frequency.  We find below, that at
the innermost stable circular orbit, $M_{\rm irr}\Omega_0\sim 0.11$.
From Fig.~\ref{fig:Spinning1}, we find for this angular frequency,
$E_{rad}/E_{\mbox{\tiny ADM}}\approx 2\cdot 10^{-4}$.  This indicates
that, when conformal flatness is assumed, we should not expect the
fractional error induced in $E_{\mbox{\tiny ADM}}$ for systems with
non-vanishing angular momentum to be larger than $\sim10^{-3}$.
Furthermore, at $M_{\rm irr}\Omega_r=0.11$, $\Delta\Omega\approx 0.0025$.
From this we expect that the rotational state of each black hole,
within the binary black hole configurations below, should deviate by at
most one or two percent from the intended values for corotating or
irrotational holes.
In Fig.~\ref{fig:Spinning1} we also plot some time-derivatives
assuming the initial data are evolved with the constructed gauge
$\Lapse, \Shift^i$.  These time-derivatives are proportional to
$(M_{\rm irr}\Omega_r)^2$.  However, their interpretation is more
difficult, due to their gauge dependence and the difficulty of finding
a meaningful normalization.

\subsection{Single boosted black holes}

A boosted single black hole in a comoving coordinate system appears
time-independent.  A well-known example is the boosted Kerr--Schild
form of a Kerr black hole.  In such comoving coordinates, the shift
does not vanish at infinity, but approaches the boost velocity of the
black hole,
\begin{equation}\label{eq:shift-BC-infty}
\Shift^i\big|_{r\to\infty} = v^i.
\end{equation}

\begin{figure}
\centerline{\includegraphics[scale=0.465]{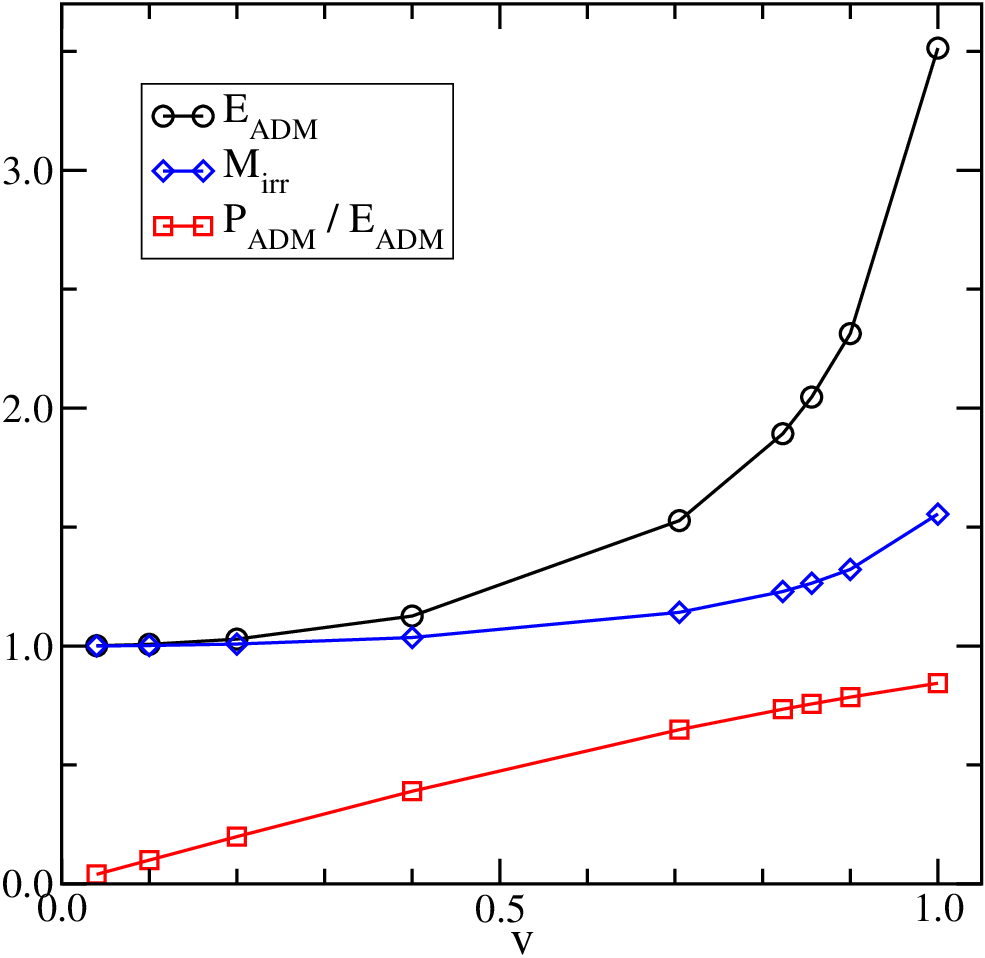}}
\caption{\label{fig:Boost1}Single boosted black hole initial-data
sets}
\end{figure}

\begin{figure}
\centerline{\includegraphics[scale=0.465]{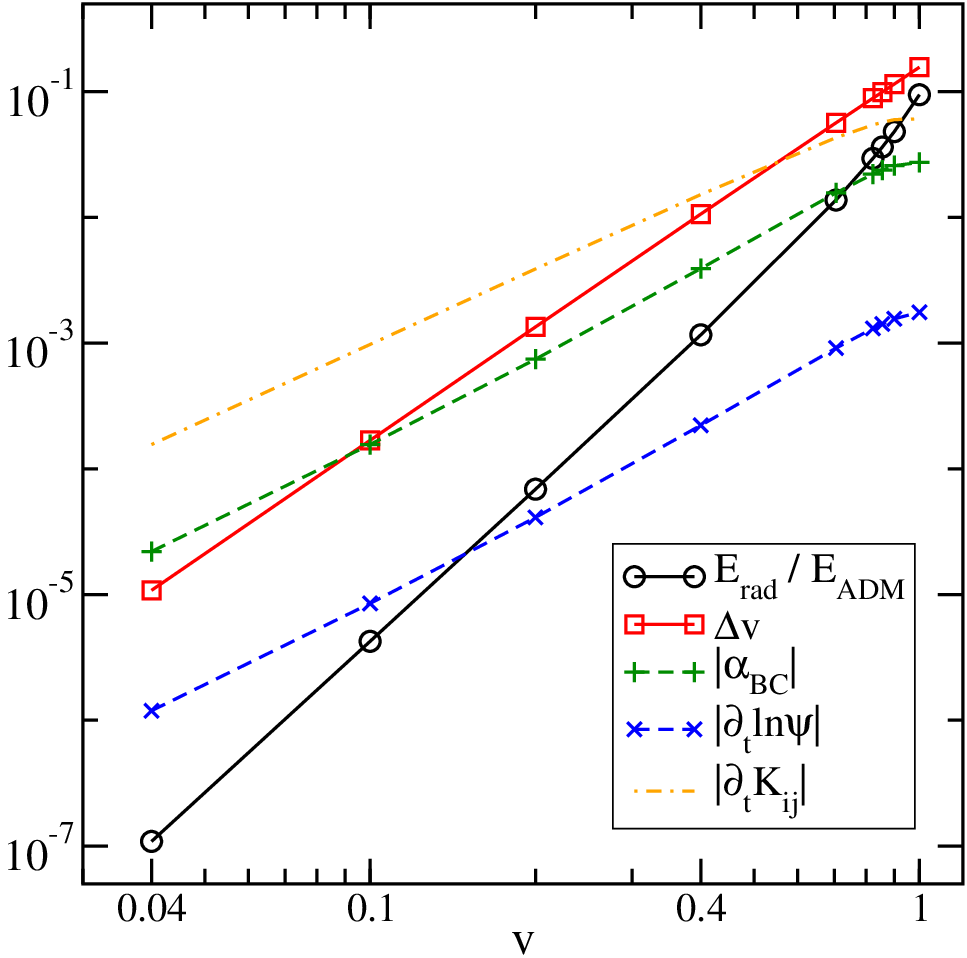}}
\caption{\label{fig:Boost2}Single boosted black hole initial-data
sets: Deviations from exact time-independence.}
\end{figure}

We apply now the quasiequilibrium formalism to construct boosted black
holes in comoving coordinates by using Eq.~(\ref{eq:shift-BC-infty})
as the boundary condition on the shift at the outer boundary.  At the
excised sphere, we set $\parShift^i\big|_{\cal S}=0$.  Furthermore, we
assume again conformal flatness, use Eqs.~(\ref{eq:K-1})
and~(\ref{eq:Lapse-BC-1}) to fix the mean curvature and the lapse BC,
and excise a coordinate sphere with radius $r_{exc}$.  The remaining
free parameter is the magnitude of the boost velocity, $v$.
Figure~\ref{fig:Boost1} presents the ADM-energy, irreducible mass and
$P_{\mbox{\tiny ADM}}/E_{\mbox{\tiny ADM}}$ as a function of the
boost-velocity.  $P_{\mbox{\tiny ADM}}/E_{\mbox{\tiny ADM}}$ is linear
in $v$ for small $v$, as it should be.  As $v$ approaches unity,
$E_{\mbox{\tiny ADM}}$ strongly increases.  Figure~\ref{fig:Boost2}
presents measures of how faithfully these initial-data sets represent
a boosted stationary black hole.  The maximum radiation content
$E_{rad}$ grows as $v^4$.  At the ISCO we can estimate that $v\sim0.4$
where we find $E_{rad}\approx 10^{-3}E_{\mbox{\tiny ADM}}$.  In order
to measure how well the special relativistic relation $v=P/E$ is
satisfied, we define
\begin{equation}
\Delta v \equiv v - \frac{P_{\mbox{\tiny ADM}}}{E_{\mbox{\tiny ADM}}};
\end{equation}
we find that $\Delta v\propto v^3$ for small $v$, with $\Delta
v\approx 0.01$ for $v=0.4$.

\section{Quasi-circular orbits for black-hole binaries}
\label{sec:quasi-circ-orbits}

In the case of a single black hole, if appropriate choices for the
freely specifiable data and boundary conditions are made, then an
exact equilibrium solution of the initial-data equations can be found.
However for binary black-hole configurations, no such true equilibrium
or stationary state exists.  This is a much more stringent test of the
quasiequilibrium boundary conditions.  In this section, we will
examine the solutions for the case of equal-mass black-hole binaries
that are either corotating or irrotational.

We will consider binary configurations over a range of separations.  A
black hole is represented in the coordinate system by an excised
2-surface.  The relative coordinate sizes of various excised surfaces
parameterize the relative sizes of the resulting physical black holes.
The coordinate separations of the holes parameterize their physical
separation.  These coordinate sizes and separations are measured in
the coordinates associated with the chosen conformal metric.  In this
work, all excised surfaces are the surfaces of coordinate spheres.
Furthermore, for the simple cases of corotating and irrotational
binaries, equal-mass black holes are obtained by choosing excision
surfaces for the two holes that have equal radii.

Before solving the initial-data equations, we must make choices for
the freely specifiable data. In all cases, we will make use of the
quasiequilibrium assumptions on the free data that $\CMtd_{ij}=0$ and
$\dtime\TrExCurv=0$.  We also continue to use the approximation that
the conformal three-geometry is flat (ie.\ that $\CMetric_{ij}$ is a
flat metric).  The remaining free data is the trace of the extrinsic
curvature, $\TrExCurv$.  For this, we will consider two choices:
maximal slicing with $\TrExCurv=0$, and a non-maximal slicing based on
Eddington--Finkelstein slicing.

In addition to the freely specifiable data, we must also fix the
boundary conditions on the excision surfaces that correspond to the
surface of each black hole and at the outer boundary of the
computational domain.  The outer boundary conditions were discussed at
the end of Sec.~\ref{sec:conf-thin-sandw}.  The boundary condition on
the shift, Eq.~(\ref{eq:shift_BC_infty}), contains a free parameter,
$\Omega_0$, that determines the orbital angular velocity of the
system.  The value of this parameter is chosen by demanding that the
ADM and Komar masses of the system must be
equal\cite{gourgoulhon-etal-2002a,gourgoulhon-etal-2002b,Cook-2002}.
This is a quasiequilibrium condition that is satisfied by a single
value of $\Omega_0$ and places the binary in a nearly circular orbit.

For the excision boundaries, we will use the apparent horizon
condition given by Eq.~(\ref{eq:AH_BC}) as a boundary condition of the
conformal factor $\CF$, and we will use Eq.~(\ref{eq:perpShift_BC}) to
fix the component of the shift that is normal to the excision
surface. Boundary conditions on the components of the shift that are
tangent to the excision surface depend on our choice for the spins of
the black holes.  For the case of corotation, we will demand that
$\parShift^i=0$.  However, the irrotational case requires a somewhat
more complicated choice.

The condition of quasiequilibrium requires that we choose the
tangential components of the shift so that they have the form
given in Eq.~(\ref{eq:Killing_shift_def}).  For irrotational
black holes in a binary, it is reasonable to choose the conformal
Killing vector $\xi^i$ so that it represents rotation about an axis
that is orthogonal to the plane of the orbit.  If we let $\Omega_r^i$
represent an angular velocity vector that is orthogonal to the
plane of the orbit, and if we use Cartesian coordinates for our
flat conformal metric, then Eq.~(\ref{eq:Killing_shift_def}) can
be written as
\begin{equation}
\parShift^i = \Omega_r^j x^k_\pm \epsilon^{ijk},
\label{eq:irrotBC}
\end{equation}
where $x^i_\pm \equiv x^i - C^i_\pm$ and $C^i_\pm$ is the Cartesian
coordinate location of the center of either of two excision spheres.
Finally, we take the magnitude of $\Omega_r^i$ to be equal to the
orbital angular velocity of the binary system as measured at infinity,
$\Omega_0$.  We note that there is no rigorous proof that these
choices lead to an irrotational binary system.  However, as argued in
Sec.~\ref{sec:excis-bound-cond}, especially the paragraphs leading to
Eq.~(\ref{eq:parShift-nonrotating}), these choices seem reasonable.

Finally, we must choose a boundary condition on the lapse at the
excision boundaries.  For all choices of the black-hole spins and
choices for $\TrExCurv$, we repeat the computations for three
different lapse boundary conditions, namely
Eqs.~(\ref{eq:Lapse-BC-1})---(\ref{eq:Lapse-BC-3}).

\begin{figure}
\includegraphics[scale=0.465]{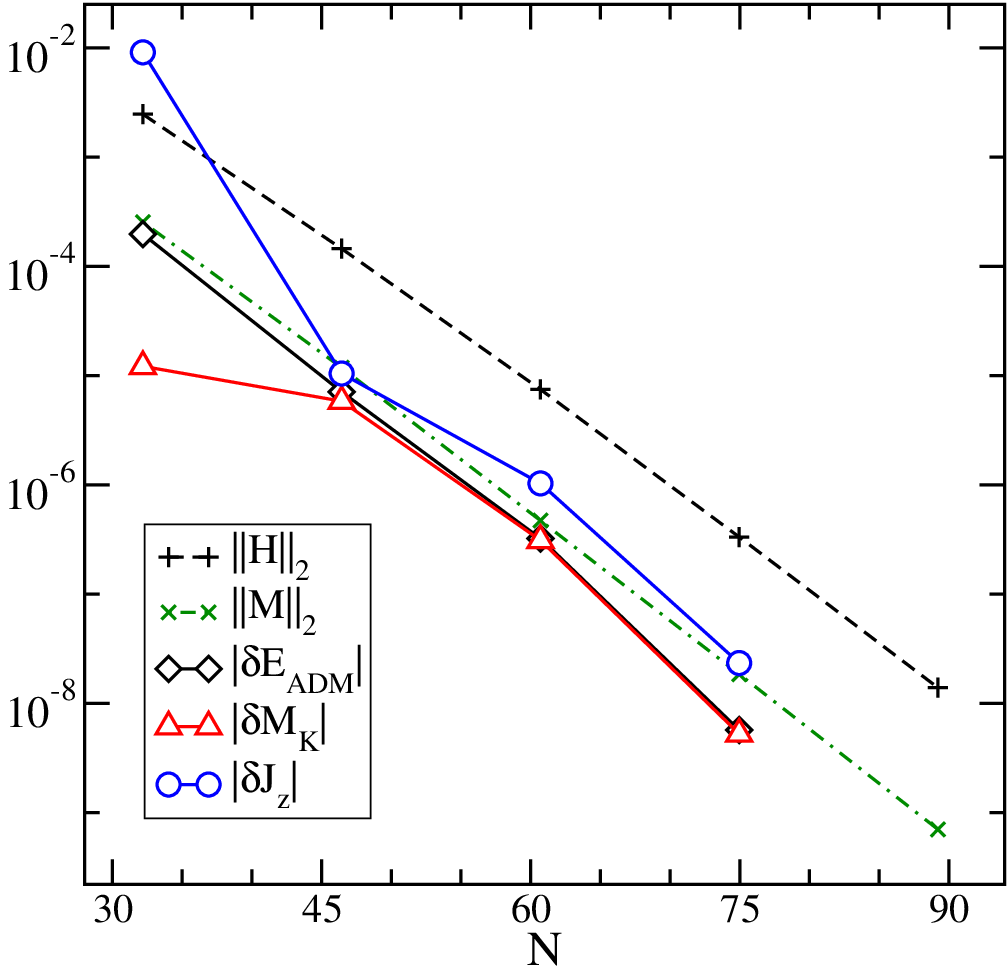}
\caption{\label{Fig:ConvergenceBBH}Convergence of the elliptic solver
for the binary black hole configurations.  Plotted are the constraint
violations in Hamiltonian and momentum constraint, and differences to
the highest resolution solve.  $N$ is the cube-root of the total
number of grid-points.}
\end{figure}

The conformal thin-sandwich equations are solved with the
pseudo-spectral collocation method described in \cite{Pfeiffer2003}.
The computational domain consists of one inner spherical shell around
each excised sphere, which overlap 43 rectangular blocks, which in
turn overlap an outer spherical shell extending to $r_{out}=10^9$.
Fig.~\ref{Fig:ConvergenceBBH} shows convergence of this solver with
spatial resolution for one typical configuration (separation $d=9$,
$K=0$, corotating black holes).  The calculations below are performed at 
a resolution comparable to $N=60$, so that the discretization errors in
$E_{\mbox{\tiny ADM}}$ and $M_{\mbox{\tiny K}}$ should be about $10^{-6}$.

\subsection{Maximal slicing}
\label{sec:maximal-slicing}

\subsubsection{Corotating binary systems}
\label{sec:corot-sys-MS}

\begin{figure}[!htbp]
\includegraphics[scale=0.465]{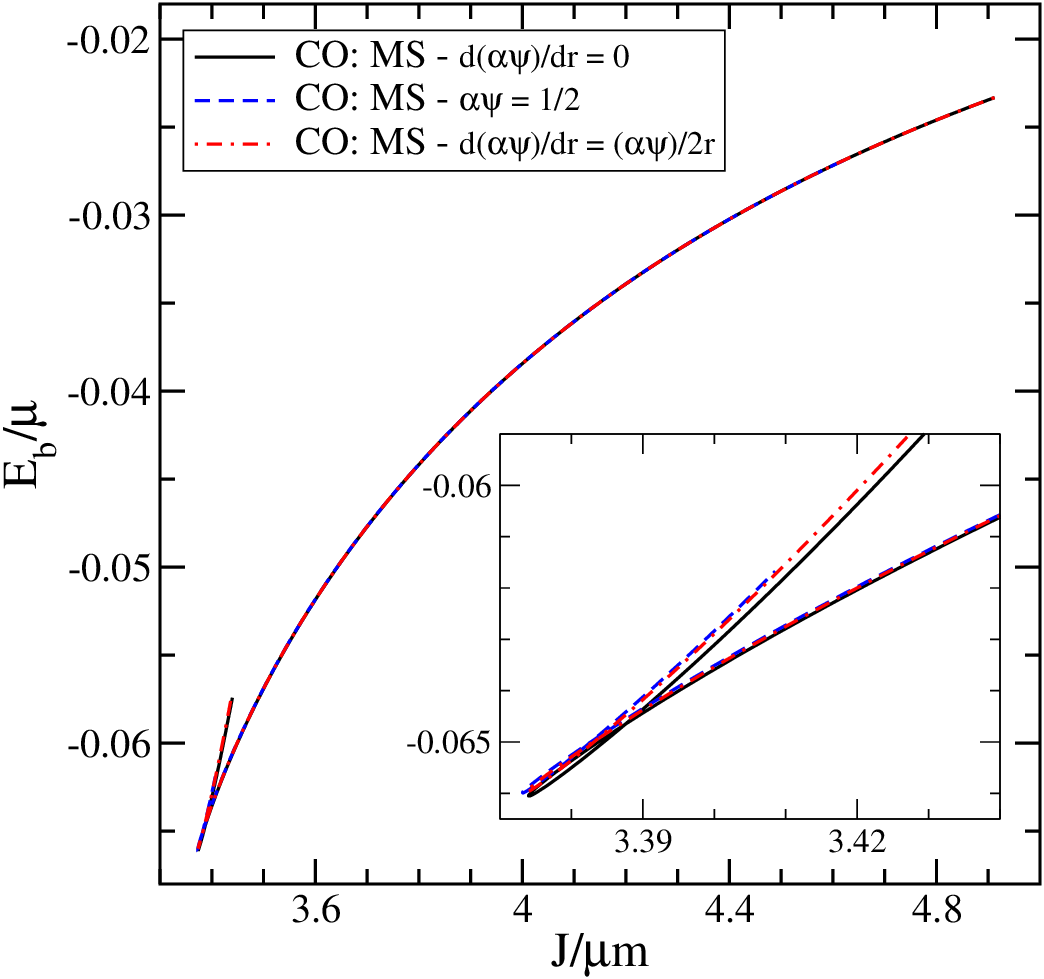}
\caption{Constant $M_{\rm irr}$ sequence of {\em corotating} equal-mass
black holes.  Maximal slicing is used in these cases, and three
different excision boundary conditions for the lapse were used.}
\label{fig:Eb_J_MSCO}
\end{figure}

We now compute initial-data sets corresponding to a binary black 
hole system in a quasi-circular orbit for many different separations.
Figure~\ref{fig:Eb_J_MSCO} shows the binding energy $E_b$ of the
binary system as a function of the total angular momentum of the
system for all the lapse boundary conditions.  The binding energy is
defined as $E_b \equiv E_{\mbox{\tiny ADM}} - m$ where $E_{\mbox{\tiny ADM}}$ is the total (ADM)
energy of the system and $m = m_1 + m_2$ is the total mass of the
system.  For the quasiequilibrium numerical results described in this
paper, we take $m_{1|2} \equiv \sqrt{\frac{A_{1|2}}{16\pi}}$ as the
irreducible mass of each individual black hole and $A_{1|2}$ are the
areas of the apparent horizon of each hole.  The reduced mass is
defined as $\mu \equiv \frac{m_1 m_2}{m}$.

We note that the choice of the lapse boundary condition has very little
effect on the solutions.  This is consistent with out assertion that
the choice of the lapse boundary condition is part of the initial
temporal gauge choice.  The inset in Fig.~\ref{fig:Eb_J_MSCO} shows
a magnified view of the region where the black holes are closest to
each other.  Even in this region, the result of the different lapse
boundary conditions are nearly indistinguishable.  Because of this,
subsequent plots displaying corotating maximal slicing results will
only display one of these sequences.

\begin{figure}[!htbp]
\includegraphics[scale=0.465]{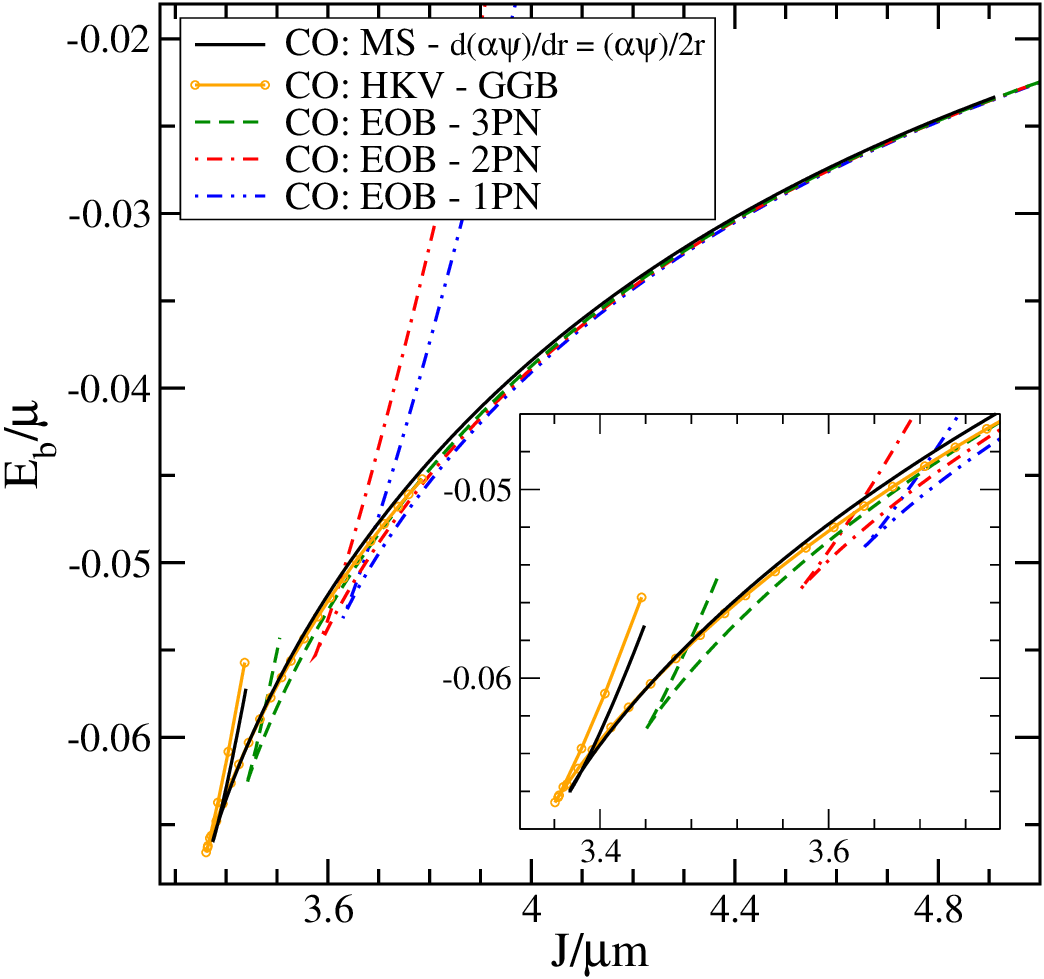}
\caption{Constant $M_{\rm irr}$ sequence of {\em corotating} equal-mass
black holes.  Comparison of post-Newtonian EOB sequences with
numerical maximal slicing results from HKV and this paper.}
\label{fig:Eb_J_COCMP}
\end{figure}

Figure~\ref{fig:Eb_J_COCMP} shows a comparison of the same data to
analogous results obtained by Grandcl\'ement et
al.\cite{gourgoulhon-etal-2002b} (labeled CO:HKV-GGB in figure
legends), and effective one-body post-Newtonian results as reported in
Ref.~\cite{Damour-etal-2002}.  First, second, and third post-Newtonian (PN)
results are displayed (labeled CO:EOB-1PN, CO:EOB-2PN, and CO:EOB-3PN
respectively in figure legends).  The 3PN results correspond to the 
approach labeled ``3PN corot.\ $\bar{A}(u,\hat{a}^2)$'' in Table~I of
Ref.~\cite{Damour-etal-2002}.

There is good agreement between all of the results at large
separation.  Also, it appears that the PN results are converging
toward the quasiequilibrium numerical results, even when the black
holes are quite close to each other as seen in the figure's inset.  We
also see that the numerical results obtained by Grandcl\'ement et
al.\cite{gourgoulhon-etal-2002b} (hereafter GGB) differ only slightly
from the quasiequilibrium results.  As discussed in
Ref.~\cite{Cook-2002}, the numerical solutions obtained by GGB must
violate the constraints.  The agreement seen in
Fig.~\ref{fig:Eb_J_COCMP} lends support to the belief that the
violation of the constraints is, in some sense, small and has a
small impact on the physical content of the data.

Another method for comparing data for the circular orbits of compact
binaries is to examine the location of the inner-most stable circular
orbit (ISCO).  The ISCO is not a well defined concept in general.
However, in situations where the dissipative effects of radiation
reaction have been eliminated, an ISCO becomes more meaningful.  The
ISCO is defined in terms of a minimum of some appropriate energy.  For
corotating binary systems, true stationary configurations can exist
(although they contain an infinite amount of energy in the form of
gravitational radiation, see Ref.~\cite{friedman-etal-2002}).  In this
case, the minimum of the total energy can be rigorously associated
with the onset of a secular instability\cite{sorkin81,sorkin82}.  In
the absence of true stationary configurations, this ``turning point''
method is still used to define the ISCO.

In order to locate a turning point, one must have a sequence of binary
configurations spanning a range of separations.  How this sequence is
constructed is not uniquely defined.  The ambiguity arises because of
the lack of a fixed fundamental length scale in the problem.  In the
first work to construct sequences of black-hole binary initial-data
sets representing circular orbits\cite{cook94e}, the total mass of the
black hole (defined in terms of the Christodoulou mass
formula\cite{christ70}) was used to normalize the sequences.  

GGB suggest another approach based on the thermodynamic
identity\cite{friedman-etal-2002}
\begin{equation}
  \td{E_{\mbox{\tiny ADM}}} = \Omega_0\, \td{J_{\mbox{\tiny ADM}}}.
\label{eq:thermo_identity}
\end{equation}
This identity should be satisfied by a true stationary sequence of
corotating black holes.  Let $s$ denote some parameter along a
sequence of initial-data sets, and let $e(s)$, $j(s)$, and $\omega(s)$
denote the numerical values for dimensionless versions of the total
energy, total angular momentum, and orbital angular velocity at
location $s$ along the sequence.  We are free to define a fundamental
length scale $\chi(s)$ along the sequence in any way we like, so long
as we define the dimensionful total energy $E_{\mbox{\tiny ADM}}(s)$, total angular
momentum $J_{\mbox{\tiny ADM}}(s)$, and orbital angular velocity $\Omega_0(s)$
consistently via,
\begin{eqnarray}
E_{\mbox{\tiny ADM}}(s) &\equiv& \chi(s)e(s),
\label{eq:energy_scale} \\
J_{\mbox{\tiny ADM}}(s) &\equiv& \chi^2\!(s)j(s),
\label{eq:angmom_scale} \\
\Omega_0(s) &\equiv& \chi^{-1}\!(s)\omega(s).
\label{eq:angvel_scale}
\end{eqnarray}
Enforcing the identity (\ref{eq:thermo_identity}) is sufficient to
determine the change in $\chi(s)$ between two points on the sequence.
If we integrate along the sequence from a point $s_1$ to another
point $s_2$, then we find that
\begin{equation}
  \chi(s_2) = \chi(s_1)\exp\left\{-\int_{s_1}^{s_2}{
    \frac{e^\prime(s) - \omega(s)j^\prime(s)}{e(s) - \omega(s)j(s)}}ds\right\},
\label{eq:thermo_scale}
\end{equation}
where a prime denotes differentiation along the sequence.

\begin{figure}[!htbp]
\includegraphics[scale=0.465]{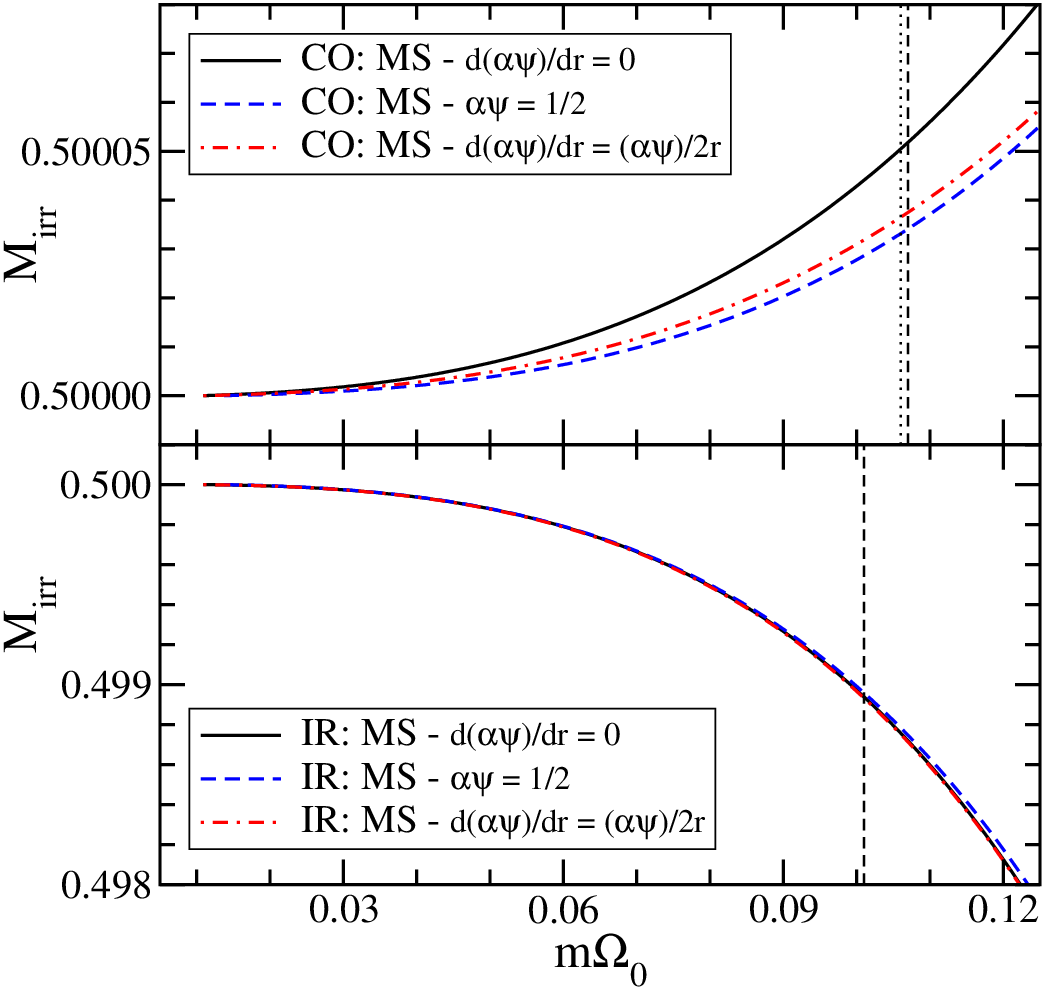}
\caption{Plot of the irreducible mass of one black hole when the
sequence of solutions is normalized to maintain $\td{E}=\Omega\td{J}$.
Three different choices for the lapse boundary condition are shown for
both corotating (upper plot) and irrotational (lower plot) black holes
in an equal-mass binary.  The vertical dashed line shows the
approximate location of the ISCO defined by a minimum in $E_b$.  The
vertical dotted line shows the approximate location of the ISCO
defined by a minimum in $E_{\mbox{\tiny ADM}}$.  In the irrotational
case, this latter ISCO line is off the plot to the right.}
\label{fig:Mirr_O}
\end{figure}

If the sequence is normalized via Eq.~(\ref{eq:thermo_identity}), then
the irreducible mass of one black hole, $M_{\rm irr}=\frac12m$, is not
necessarily constant along the sequence.  The top half of
Fig.~\ref{fig:Mirr_O} shows $M_{\rm irr}$ for a corotating
quasiequilibrium equal-mass binary as a function of the orbital
angular velocity.  The length scale has been normalized so that
$M_{\rm irr}=1/2$ at infinite separation.  We confirm the finding of
GGB that $M_{\rm irr}$ is nearly constant along the sequence.  While
there is a clear increase in the mass as the separation decreases,
this increase is small and appears to be of roughly the same order of
magnitude as the differences due to using different lapse boundary
conditions.  As we will see later, this behavior is not mirrored in
the irrotational data.

Yet another approach for normalizing the sequence is to demand that
the individual irreducible masses associated with the apparent horizons
remain constant.  This is a particularly convenient normalization for
numerical work since it relies on a well defined and easily measured
geometric quantity.  Each of these choices for normalizing an initial
data sequence can effect the location of the ISCO, so it is important
to use a consistent definition when comparing data.

It is also important to clearly define which energy is being
extremised when using a turning-point method to locate the ISCO.  We
can consider using the minimum in either the ADM energy $E_{\mbox{\tiny ADM}}$ or
the binding energy $E_b$ along any sequence to define the ISCO.  From
the definition of the binding energy as $E_b \equiv E_{\mbox{\tiny ADM}} - m$, we
see that the minima will not necessarily agree if $m$ varies along
the sequence.

For the PN sequences, the ISCO is defined as the minimum in the
binding energy $E_b$ along sequences where the irreducible masses of
the black holes remain fixed.  For the PN sequences,
Eq.~(\ref{eq:thermo_identity}) is identically satisfied as well, so
this is equivalent to finding the minimum in $E_{\mbox{\tiny ADM}}$.
This is not true for the quasiequilibrium numerical data.

\begin{table}[!htbp]
\caption{Parameters of the ISCO configuration for {\em corotating}
equal-mass black holes computed with the maximal slicing condition.
Results are given for three different choices of the lapse boundary
condition and two choices for the definition of the location of the
ISCO.  For comparison, the lower part of the table lists results of
Refs.~\cite{gourgoulhon-etal-2002b, Damour-etal-2002,Blanchet:2002};
``PN standard''~\cite{Blanchet:2002} represents a post-Newtonian
expansion in the standard form without use of the
EOB-technique.}
\begin{ruledtabular}
\begin{tabular}{rc|ccc}
Lapse BC & ISCO min. 
         & $m\Omega_0$ & $E_b/m$ & $J/m^2$ \\
\hline
$\frac{\td{(\Lapse\CF)}}{\td{r}}=0$ 
         &  ADM  & 0.105 & -0.0165 & 0.844 \\
         & $E_b$ & 0.107 & -0.0165 & 0.844 \\
\hline
$\Lapse\CF=\frac12$ 
         &  ADM  & 0.106 & -0.0165 & 0.843 \\
         & $E_b$ & 0.107 & -0.0165 & 0.843 \\
\hline
$\frac{\td{(\Lapse\CF)}}{\td{r}}= \frac{\Lapse\CF}{2r} $ 
         &  ADM  & 0.106 & -0.0165 & 0.843 \\
         & $E_b$ & 0.107 & -0.0165 & 0.843 \\
\hline
\hline
\multicolumn{2}{@{\hspace*{1cm}}l|}{HKV-GGB}
                 & 0.103 & -0.017 & 0.839 \\
\hline
\multicolumn{2}{@{\hspace*{1cm}}l|}{1PN EOB}
                 & 0.0667 & -0.0133 & 0.907 \\
\multicolumn{2}{@{\hspace*{1cm}}l|}{2PN EOB}
                 & 0.0715 & -0.0138 & 0.893 \\
\multicolumn{2}{@{\hspace*{1cm}}l|}{3PN EOB}
                 & 0.0979 & -0.0157 & 0.860 \\
\hline
\multicolumn{2}{@{\hspace*{1cm}}l|}{1PN standard}
                 & 0.5224 & -0.0405 & 0.621 \\
\multicolumn{2}{@{\hspace*{1cm}}l|}{2PN standard}
                 & 0.0809 & -0.0145 & 0.882 \\
\multicolumn{2}{@{\hspace*{1cm}}l|}{3PN standard}
                 & 0.0915 & -0.0153 & 0.867
\end{tabular}
\end{ruledtabular}
\label{tab:ISCO_MSCO}
\end{table}

Table~\ref{tab:ISCO_MSCO} displays the dimensionless orbital angular
velocity, binding energy, and total angular momentum of the ISCO for
corotating equal-mass black holes on a maximal slice.  For the
quasiequilibrium data defined in this paper, two definitions of the
ISCO are listed.  One uses the minimum in $E_{\mbox{\tiny ADM}}$ along
sequences where condition (\ref{eq:thermo_identity}) is satisfied to
define the ISCO.  The alternative method uses a minimum in $E_b$ along
sequences where $M_{\rm irr}$ is held fixed.  However, we note that
the minima in $E_b$ along sequences that satisfy
(\ref{eq:thermo_identity}) are numerically indistinguishable from the
latter.  The ISCO for the GGB data (listed as HKV-GGB in the table) is
defined as the minima in $E_b$ along a sequence where $M_{\rm irr}$
remains fixed.  Recall that the definition of the ISCO for the PN data
is consistent with {\em either} definition used for the
quasiequilibrium data.  In addition to the effective one-body (EOB) PN
data displayed previously in Fig.~\ref{fig:Eb_J_COCMP}, we also
include ``standard'' PN results for the ISCO as reported in
Ref.~\cite{Blanchet:2002}.

For corotating quasiequilibrium data, there is very little difference
in the results for the two definitions of the ISCO.  As we will see
later, this is {\em not} true for the irrotational configurations (see
Table~\ref{tab:ISCO_MSIR}).  In that case, it is clear that only the
definition in terms of $E_b$ is consistent with the PN data.

\begin{figure}[!htbp]
\includegraphics[scale=0.465]{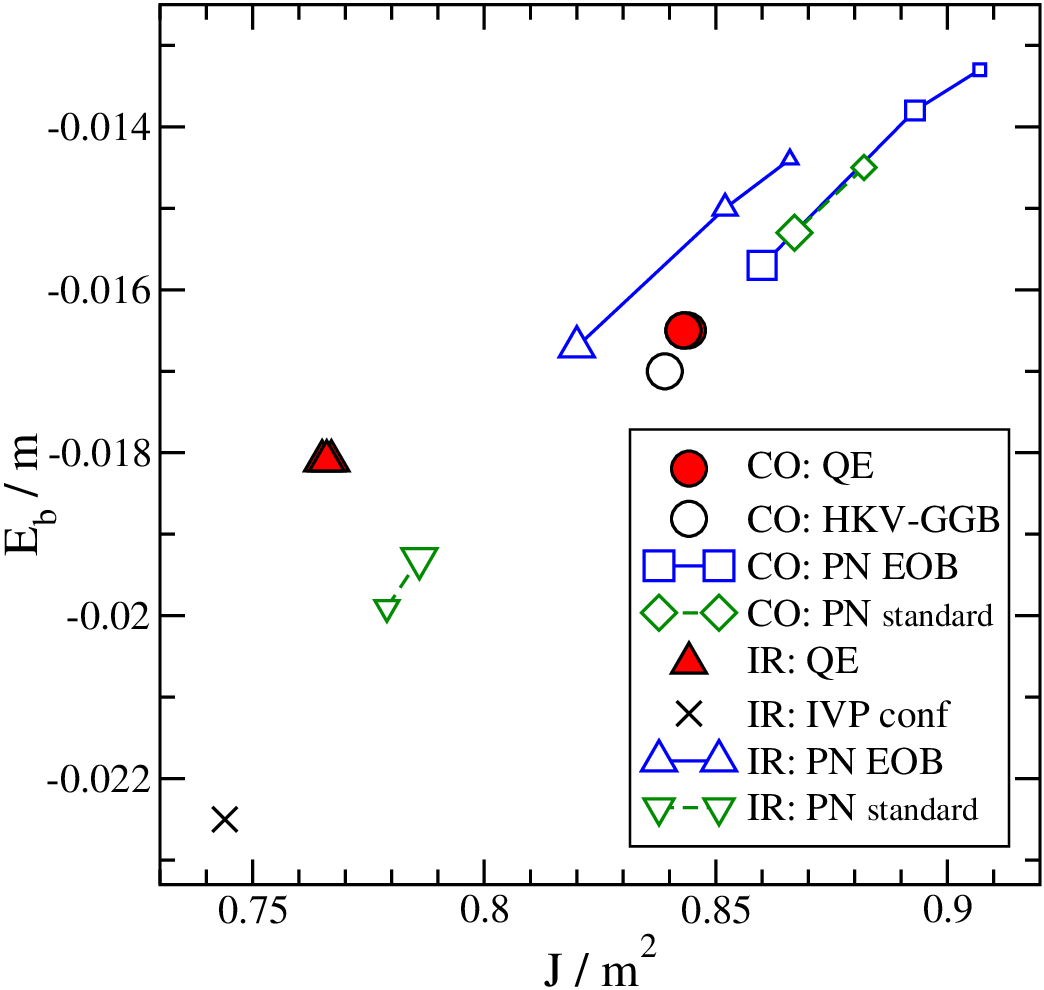}
\caption{ISCO configuration for three different choices of the lapse
boundary condition for equal-mass corotating and irrotational black
holes computed with the maximal slicing condition.  For comparison,
results of Refs.~\cite{cook94e, gourgoulhon-etal-2002b,
Damour-etal-2002,Blanchet:2002} are included. For post-Newtonian
calculations the size of the symbol indicates the order, the largest
symbol being 3PN.  ``PN standard''~\cite{Blanchet:2002} represents a
PN-expansion in the standard form without use of the
EOB-technique (only 2PN and 3PN are plotted).}
\label{fig:Eb_J_ISCO}
\end{figure}

Figure~\ref{fig:Eb_J_ISCO} plots binding energy versus orbital angular
velocity for the ISCO obtained for all three lapse boundary conditions
for the corotating quasiequilibrium data, as well as the corotating
results from GGB and PN results.  All the numerical results are
computed on a maximal slice.  We see that the results for the
different lapse boundary conditions are essentially indistinguishable.
We also see that the PN results converge roughly toward the numerical
quasiequilibrium results.  While we would not expect the
quasiequilibrium numerical results to agree with any of the individual
PN results, we might expect the GGB result to agree within numerical
error.  All of our numerical results using different lapse boundary
conditions are essentially indistinguishable.  Furthermore, if we use
a lapse boundary condition that approaches a Dirichlet value of zero
on the excision surface, the resulting set of boundary conditions is
equivalent to those used by GGB.  The difference seen between the GGB
ISCO and our results may well be due to the regularization procedure
introduced by GGB.

\begin{figure}[!htbp]
\includegraphics[scale=0.465]{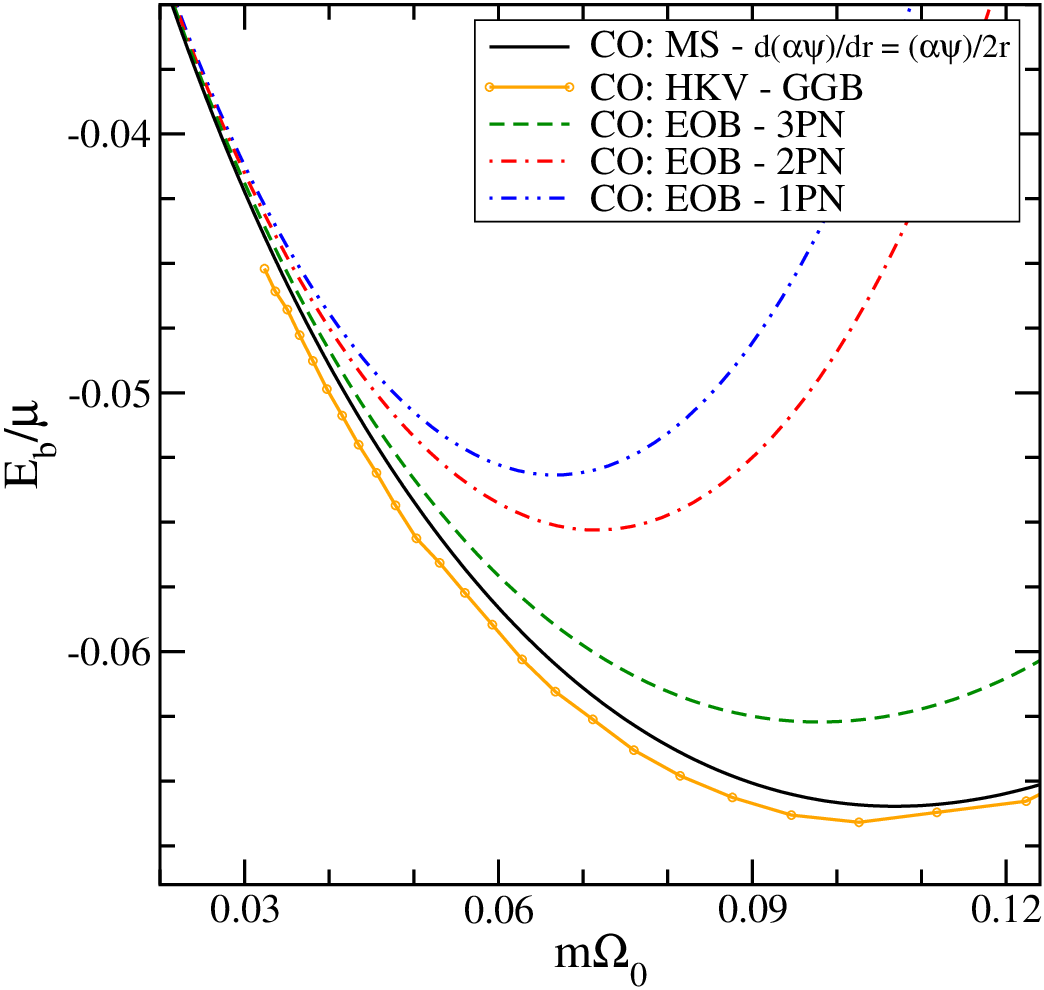}
\caption{Constant $M_{\rm irr}$ sequence of {\em corotating} equal-mass
black holes.  Comparison of post-Newtonian EOB sequences with
numerical maximal slicing results from HKV and this paper.}
\label{fig:Eb_O_COCMP}
\end{figure}

\begin{figure}[!htbp]
\includegraphics[scale=0.465]{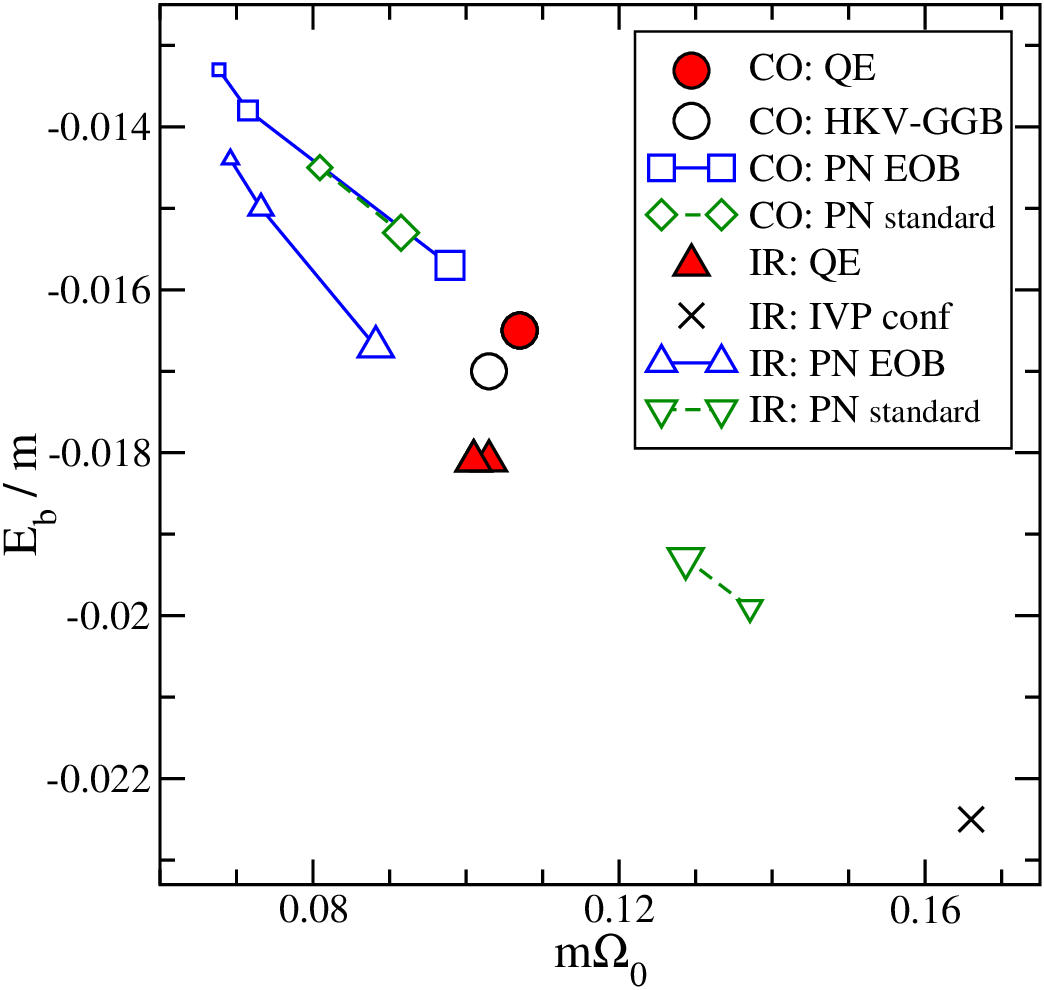}
\caption{ISCO configuration for three different choices of the lapse
boundary condition for equal-mass corotating and irrotational black
holes computed with the maximal slicing condition.  Symbols as in
Fig.~\ref{fig:Eb_J_ISCO}.}
\label{fig:Eb_O_ISCO}
\end{figure}

\begin{figure}[!htbp]
\includegraphics[scale=0.465]{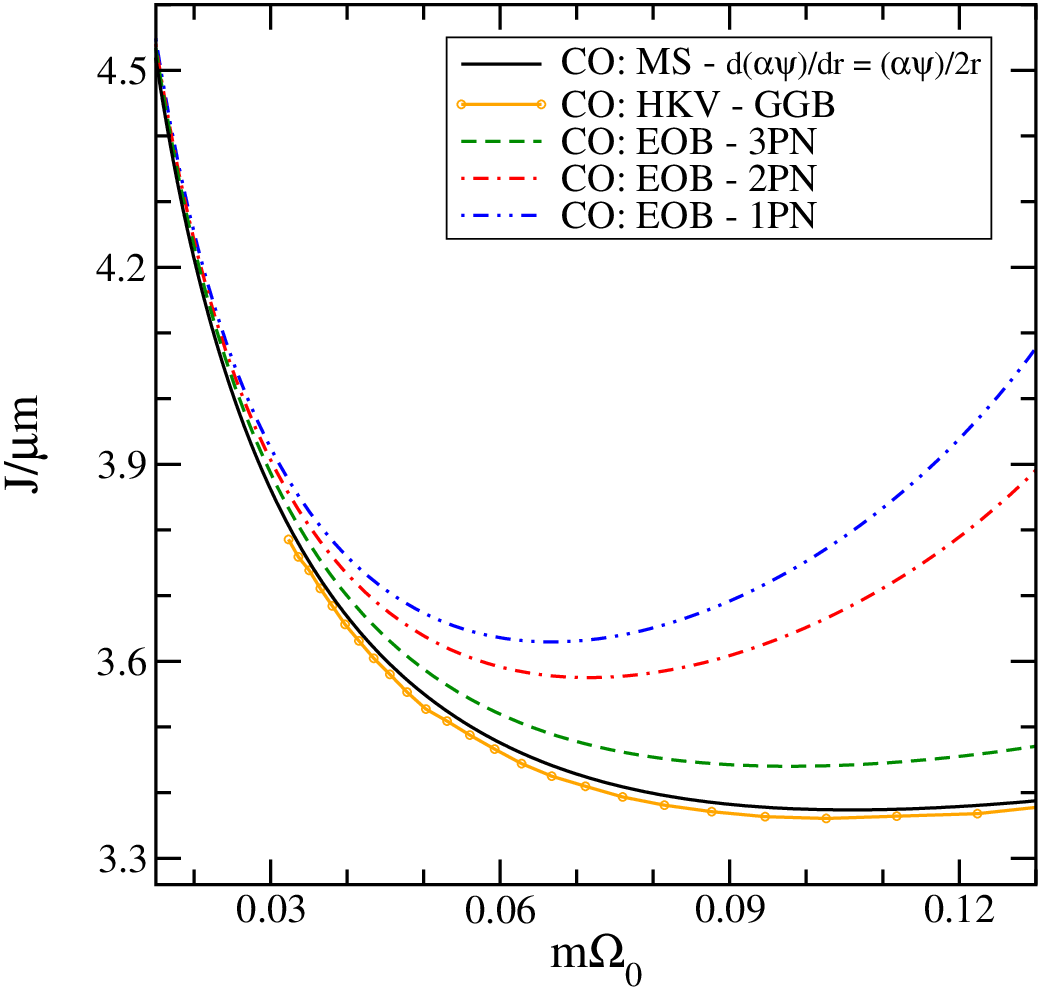}
\caption{Constant $M_{\rm irr}$ sequence of {\em corotating} equal-mass
black holes.  Comparison of post-Newtonian EOB sequences with
numerical maximal slicing results from HKV and this paper.}
\label{fig:J_O_COCMP}
\end{figure}

\begin{figure}[!htbp]
\includegraphics[scale=0.465]{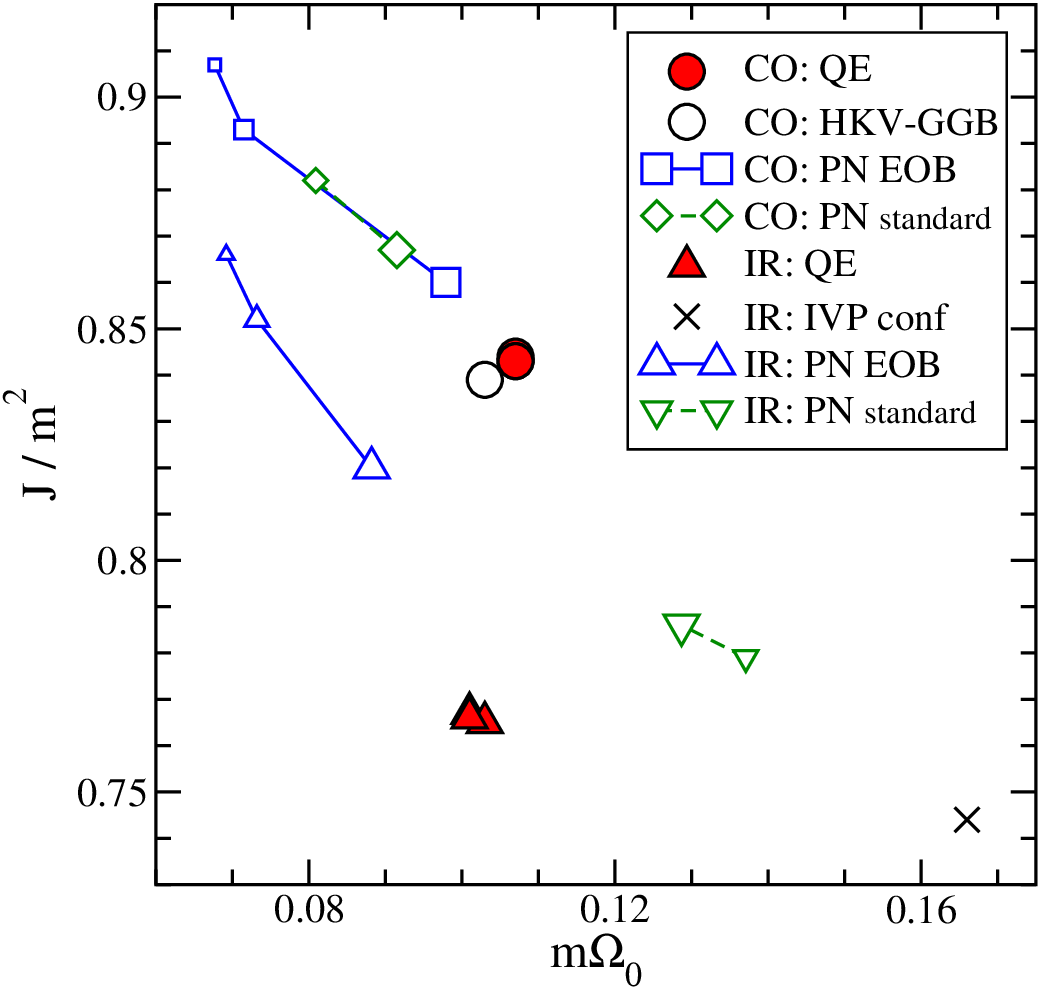}
\caption{ ISCO configuration for three different choices of the lapse
boundary condition for equal-mass corotating and irrotational black
holes computed with the maximal slicing condition.  Symbols as in
Fig.~\ref{fig:Eb_J_ISCO}.}
\label{fig:J_O_ISCO}
\end{figure}

There are three rigorously defined gauge-invariant global quantities
associated with a black-hole binary system: the ADM energy $E_{\mbox{\tiny ADM}}$,
the total angular momentum $J$, and the orbital angular velocity as
seen at infinity $\Omega_0$.  Figures~\ref{fig:Eb_J_COCMP} and
\ref{fig:Eb_J_ISCO} plotted the binding energy $E_b$ (directly related
to the ADM energy) as a function of $J$.  Figures~\ref{fig:Eb_O_COCMP}
and \ref{fig:Eb_O_ISCO} plot $E_b$ as a function of $\Omega_0$ for the
same set of sequences and for the ISCO.  And, Figs.~\ref{fig:J_O_COCMP}
and \ref{fig:J_O_ISCO} plot $J$ as a function of $\Omega_0$ for the
same set of sequences and for the ISCO.

\subsubsection{Irrotational binary systems}
\label{sec:irrot-sys-MS}

\begin{figure}[!htbp]
\includegraphics[scale=0.465]{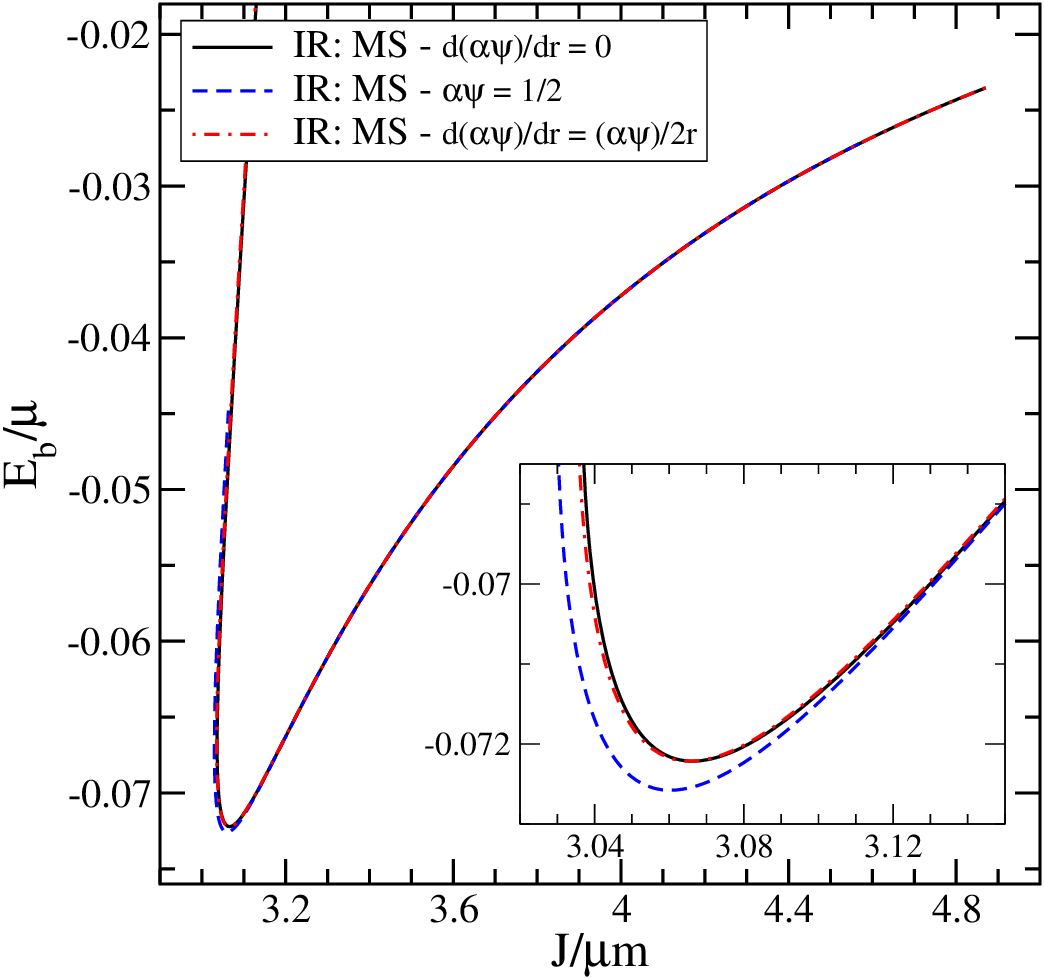}
\caption{Constant $M_{\rm irr}$ sequence of {\em irrotational} equal-mass
black holes.  Maximal slicing is used in these cases, and three
different excision boundary conditions for the lapse were used.}
\label{fig:Eb_J_MSIR}
\end{figure}

For the case of irrotational black holes, Fig.~\ref{fig:Eb_J_MSIR}
shows the binding energy $E_b$ of the binary system as a function of
$J$ for all the lapse boundary conditions.  As with the
corotating black holes, we again see that the choice of the lapse
boundary condition has very little effect on the sequence.  In the
inset to the figure we see that the effect is largest at small
separations and that the differences due to varying the lapse
boundary condition are somewhat larger than in the corotating 
case.

Perhaps the most striking difference between the corotating and
irrotational sequences is that the extrema in $E_b$ versus $J$ are
much less ``sharp'' in the irrotational sequences.  If we consider
sequences, either corotating or irrotational, that are normalized so
that $\td{E}_{\mbox{\tiny ADM}} = \Omega_0\td{J}$ is satisfied, then
the extremum in $E_{\mbox{\tiny ADM}}$ will necessarily coincide with
the extremum in $J$ leading to a very sharp cusp in a plot of
these quantities.  However, because $M_{\rm irr}$ is not necessarily
constant along a sequence with this normalization, the extremum in
$E_b$ will not necessarily coincide with that of $J$.  Thus, we should
certainly not expect to see a sharp cusp in either
Fig.~\ref{fig:Eb_J_MSCO} or Fig.~\ref{fig:Eb_J_MSIR}.

Another difference between the corotating and irrotational sequences
can be seen if Fig.~\ref{fig:Mirr_O}.  Here we see that, if we demand
that the thermodynamic identity (\ref{eq:thermo_identity}) be
satisfied along the sequence, then the variation in $M_{\rm irr}$ is
20 times larger in the irrotational sequences than in the corotating
sequences.  For the corotating sequences, the variation in $M_{\rm
irr}$ due to differences in the lapse boundary condition was
comparable to the average variation.  For the irrotational sequences,
the effect of different lapse boundary conditions is clearly
negligible.  Furthermore, we note that $M_{\rm irr}$ is {\em
decreasing} as the binary separation decreases.  This behavior is
unphysical, as the irreducible mass never decreases; therefore,
$M_{\rm irr}$ should also not decrease during the insipiral of a
binary black hole.

\begin{figure}[!htbp]
\includegraphics[scale=0.465]{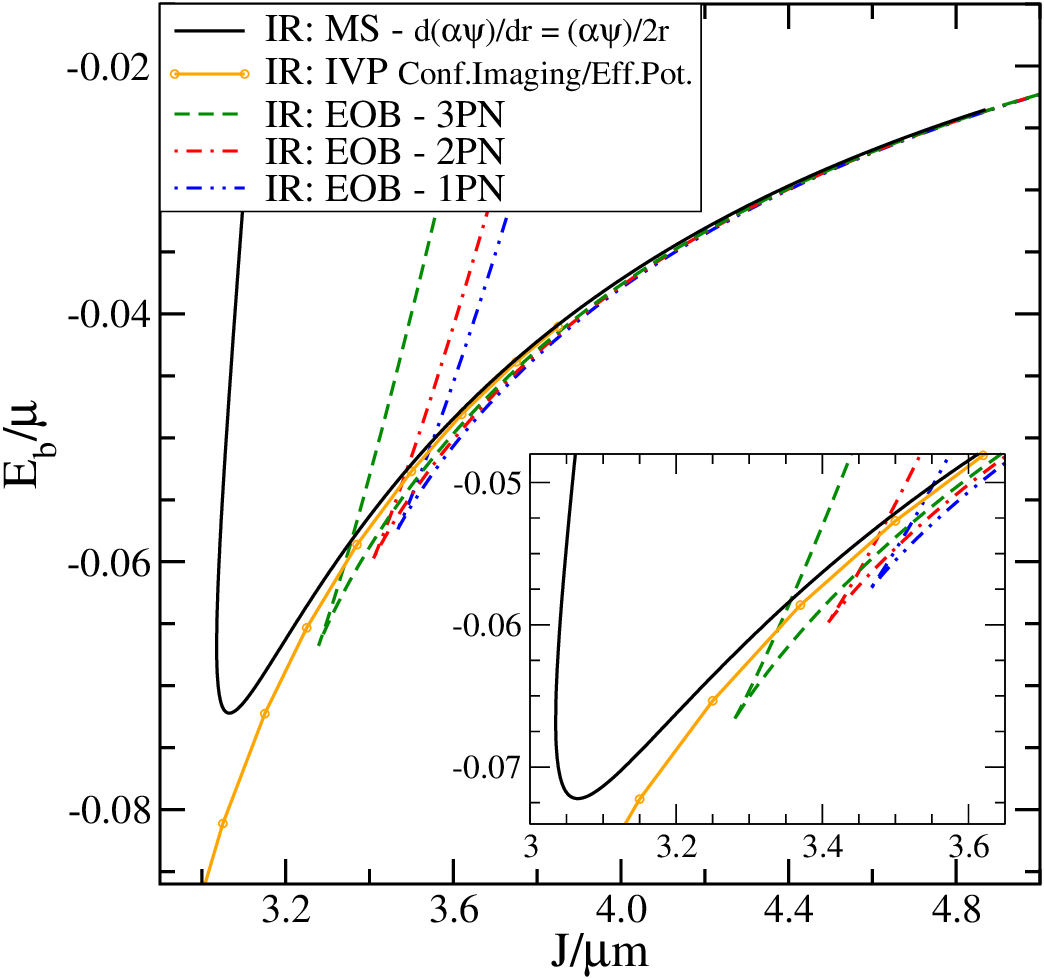}
\caption{Constant $M_{\rm irr}$ sequence of {\em irrotational} equal-mass
black holes.  Comparison of post-Newtonian EOB sequences with
numerical maximal slicing results from Conformal Image and this paper.}
\label{fig:Eb_J_IRCMP}
\end{figure}

Figure~\ref{fig:Eb_J_IRCMP} shows a comparison of our irrotational
data to the effective one-body post-Newtonian results for irrotational
holes as reported in Ref.~\cite{Damour-etal-2002}.  First, second, and
third post-Newtonian results are displayed (labeled IR:EOB-1PN,
IR:EOB-2PN, and IR:EOB-3PN respectively in figure legends).  The 3PN
results correspond to the approach labeled ``3PN corot.\
$\bar{A}(u,0)$'' in Table~I of Ref.~\cite{Damour-etal-2002}.  Also
plotted in this figure is the first sequence of numerical initial-data
solutions for an equal-mass black-hole binary in quasicircular orbit,
obtained from inversion-symmetric initial data using an effective
potential approach\cite{cook94e} (labeled IR:Conf.\ Imaging/Eff.\
Pot.\ or IVP-conf in figure legends).

Again, there is good agreement between all of the results at large
separation.  Also, it appears that the PN results are converging
toward the irrotational quasiequilibrium numerical results, even when
the black holes are quite close to each other as seen in the figure's
inset.  We also see that the early numerical results obtained from the
conformal-imaging data\cite{cook94e} differ only slightly from the
quasiequilibrium results up the location of the ISCO in the
quasiequilibrium sequence.  However, the conformal-imaging sequence
extends to much smaller separations before encountering its ISCO.

\begin{table}[!htbp]
\caption{Parameters of the ISCO configuration for {\em irrotational}
equal-mass black holes computed with the maximal slicing condition.
Results are given for three different choices of the lapse boundary
condition and two choices for the definition of the location of the
ISCO.  Layout as in Table~\ref{tab:ISCO_MSCO}.}
\begin{ruledtabular}
\begin{tabular}{rc|ccc}
Lapse BC & ISCO min. 
         & $m\Omega_0$ & $E_b/m$ & $J/m^2$ \\
\hline
$\frac{\td{(\Lapse\CF)}}{\td{r}}=0$ 
         &  ADM  & 0.144 & -0.0146 & 0.761 \\
         & $E_b$ & 0.101 & -0.0181 & 0.767 \\
\hline
$\Lapse\CF=\frac12$ 
         &  ADM  & 0.148 & -0.0145 & 0.760 \\
         & $E_b$ & 0.103 & -0.0181 & 0.765 \\
\hline
$\frac{\td{(\Lapse\CF)}}{\td{r}}= \frac{\Lapse\CF}{2r} $ 
         &  ADM  & 0.145 & -0.0146 & 0.761 \\
         & $E_b$ & 0.101 & -0.0181 & 0.766 \\
\hline
\hline
\multicolumn{2}{@{\hspace*{1cm}}l|}{Conf.~Imag.}
                 & 0.166 & -0.0225 & 0.744 \\
\hline
\multicolumn{2}{@{\hspace*{1cm}}l|}{1PN EOB}
                 & 0.0692 & -0.0144 & 0.866 \\
\multicolumn{2}{@{\hspace*{1cm}}l|}{2PN EOB}
                 & 0.0732 & -0.0150 & 0.852 \\
\multicolumn{2}{@{\hspace*{1cm}}l|}{3PN EOB}
                 & 0.0882 & -0.0167 & 0.820\\
\hline
\multicolumn{2}{@{\hspace*{1cm}}l|}{1PN standard}
                 & 0.5224 & -0.0405 & 0.621 \\
\multicolumn{2}{@{\hspace*{1cm}}l|}{2PN standard}
                 & 0.1371 & -0.0199 & 0.779 \\
\multicolumn{2}{@{\hspace*{1cm}}l|}{3PN standard}
                 & 0.1287 & -0.0193 & 0.786
\end{tabular}
\end{ruledtabular}
\label{tab:ISCO_MSIR}
\end{table}

Table~\ref{tab:ISCO_MSIR} displays the ISCO data for irrotational
equal-mass black holes on a maximal slice.  Again, the ISCO is defined
in two ways for the QE-sequence data.  However, unlike the corotating
case, the two definitions disagree dramatically.  It is clear that the
ISCO, when defined as a minimum of $E_b$ along a sequence where
$M_{\rm irr}$ is held fixed, is consistent with the results from the
PN data.  This is fortunate given the fact that $M_{\rm irr}$ {\em
decreases} as the binary separation decreases along sequences where
Eq.~(\ref{eq:thermo_identity}) is satisfied.  However, it is unclear
why these $E_{\mbox{\tiny ADM}}$ defined ISCOs disagree so
significantly from the PN data.  The $E_b$-defined ISCO data for the
various irrotational sequences is plotted with the corotating data in
Figs.~\ref{fig:Eb_J_ISCO}, \ref{fig:Eb_O_ISCO}, and
\ref{fig:J_O_ISCO}.

\begin{figure}[!htbp]
\includegraphics[scale=0.465]{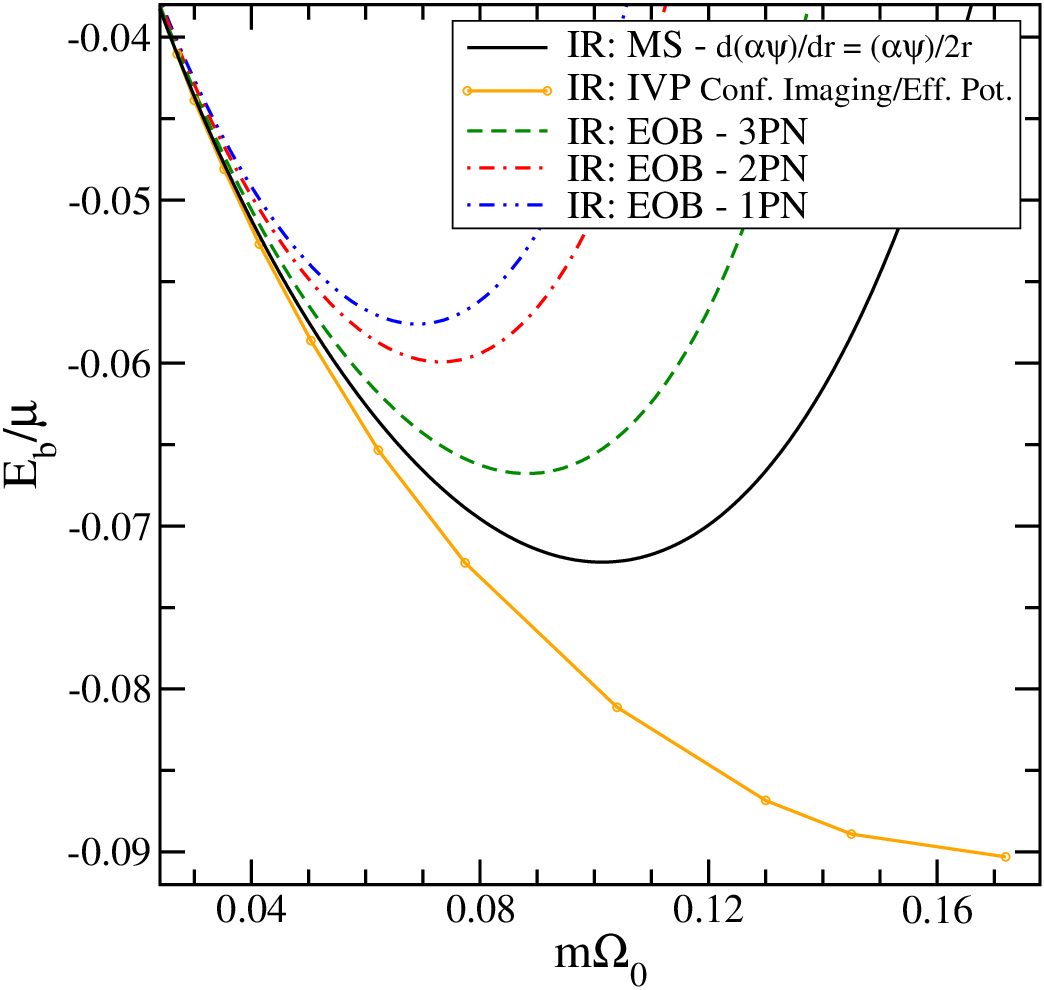}
\caption{Constant $M_{\rm irr}$ sequence of {\em irrotational} equal-mass
black holes.  Comparison of post-Newtonian EOB sequences with
numerical maximal slicing results from Conformal Image and this paper.}
\label{fig:Eb_O_IRCMP}
\end{figure}

\begin{figure}[!htbp]
\includegraphics[scale=0.465]{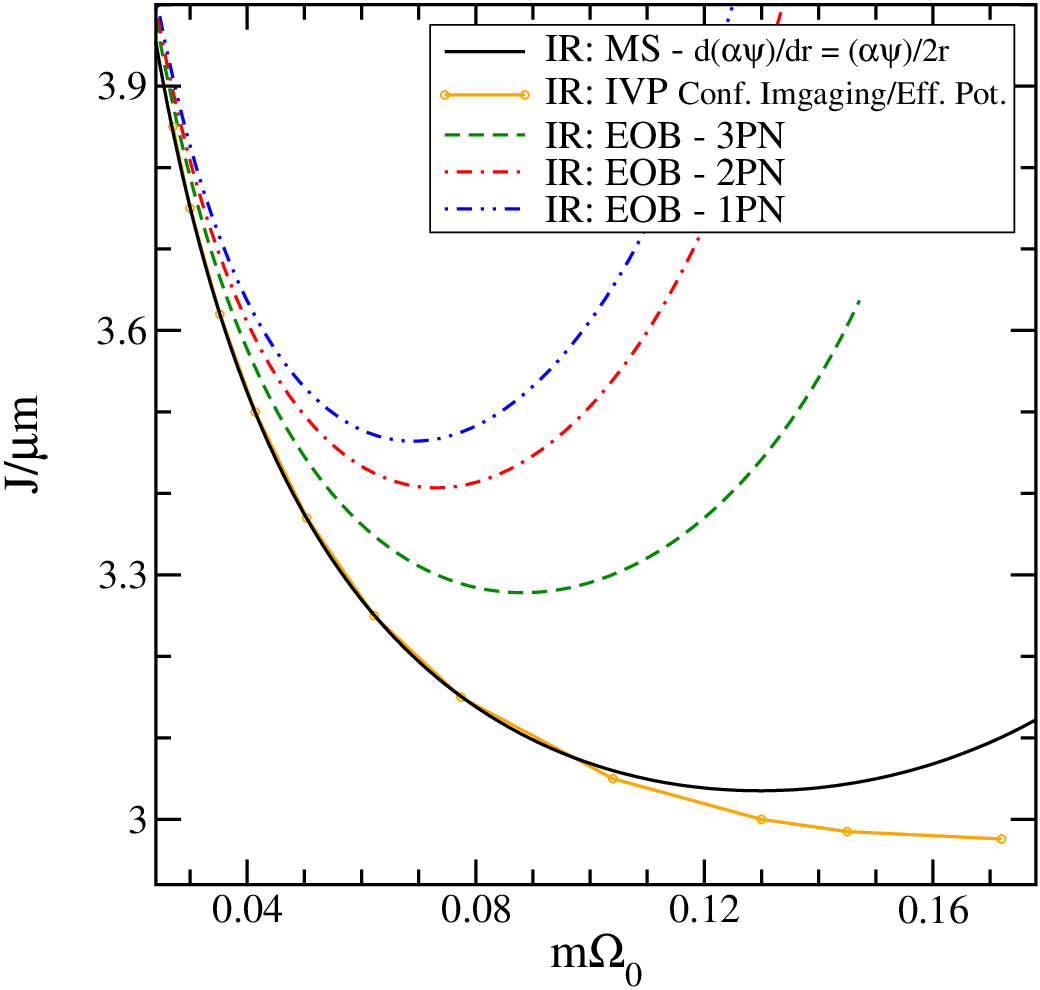}
\caption{Constant $M_{\rm irr}$ sequence of {\em irrotational} equal-mass
black holes.  Comparison of post-Newtonian EOB sequences with
numerical maximal slicing results from Conformal Image and this paper.}
\label{fig:J_O_IRCMP}
\end{figure}

Figures~\ref{fig:Eb_J_IRCMP} and \ref{fig:Eb_J_ISCO} plotted the
binding energy $E_b$ as a function of $J$ for the irrotational
sequences.  Figures~\ref{fig:Eb_O_IRCMP} and \ref{fig:Eb_O_ISCO} plot
$E_b$ as a function of $\Omega_0$ for the same set of sequences and
for the ISCO.  And, Figs.~\ref{fig:J_O_IRCMP} and \ref{fig:J_O_ISCO}
plot $J$ as a function of $\Omega_0$ for the same set of sequences and
for the ISCO.

\subsection{Non-maximal slicing}
\label{sec:non-maximal-slicing}

For the case of sequences of corotating or irrotational QE initial
data obtained on maximal slices (i.e.~$\TrExCurv\!=\!0$), the choice of the
lapse boundary condition seemed to have very little effect on gauge
invariant quantities.  This lends support to our assertion that the
choice of the lapse boundary condition is part of the initial temporal
gauge freedom.  To further test this assertion, we should consider
varying other aspects of the freely specifiable data.
Quasiequilibrium considerations demand that we choose $\CMtd_{ij}=0$
and $\dtime\TrExCurv=0$.  This leaves us the options of choosing a
non-flat conformal metric or a non-maximal slice.

While changing either can affect the content of the dynamical degrees
of freedom in the initial data, the conformal metric is more closely
tied to these dynamical degrees of freedom and to the spatial gauge
freedom.  The choice of the trace of the extrinsic curvature is
usually thought of as fixing the initial temporal gauge freedom,
although as we have seen from the example of the family of maximal
slices of Schwarzschild in Sec.~\ref{sec:bound-cond-lapse}, this is
not always sufficient to fix this aspect of the gauge freedom.  In any
case, it seems reasonable that the best choice is to vary $\TrExCurv$.

A convenient choice for a non-maximal slicing is one based on a
stationary black hole in Kerr--Schild coordinates.  For the case of a
non-charged, non-spinning black hole these are also referred to as
ingoing Eddington--Finkelstein coordinates
(cf. Ref.~\cite{kidder-etal-2000}).  The spatial metric is given by
Eq.~(\ref{eq:KS-spatial}), and after the coordinate transformation
(\ref{eq:KS-flat-r}) it becomes conformally flat.  The trace of the
extrinsic curvature for the Kerr--Schild slicing of Schwarzschild is
given by Eq.~(\ref{eq:K-3}).  For a binary system, we use a linear
combination of two copies of Eq.~(\ref{eq:K-3}), each centered at the
location of one of the black holes, to define $\TrExCurv$.

\begin{figure}[!htbp]
\includegraphics[scale=0.465]{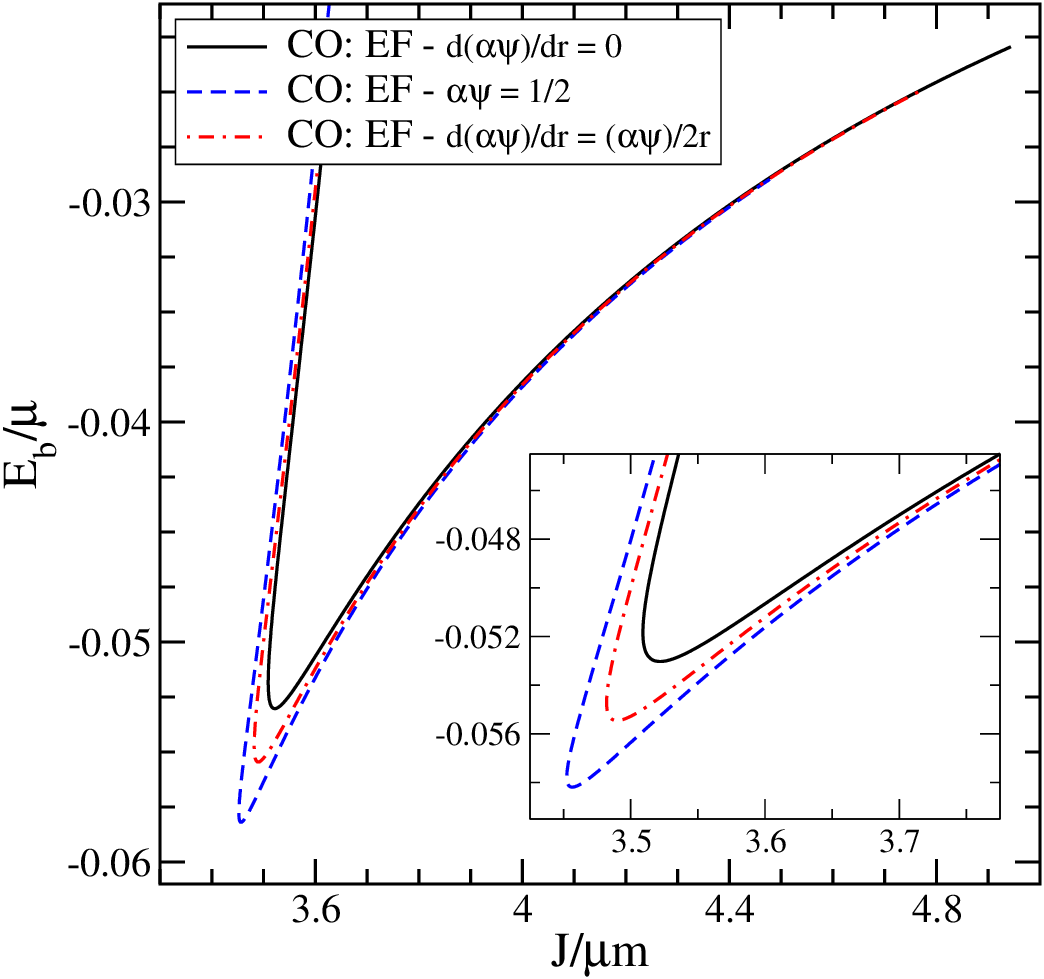}
\caption{Constant $M_{\rm irr}$ sequence of {\em corotating} equal-mass
black holes.  Eddington--Finkelstein slicing is used in these cases,
and three different excision boundary conditions for the lapse were
used.}
\label{fig:Eb_J_EFCO}
\end{figure}

\begin{figure}[!htbp]
\includegraphics[scale=0.465]{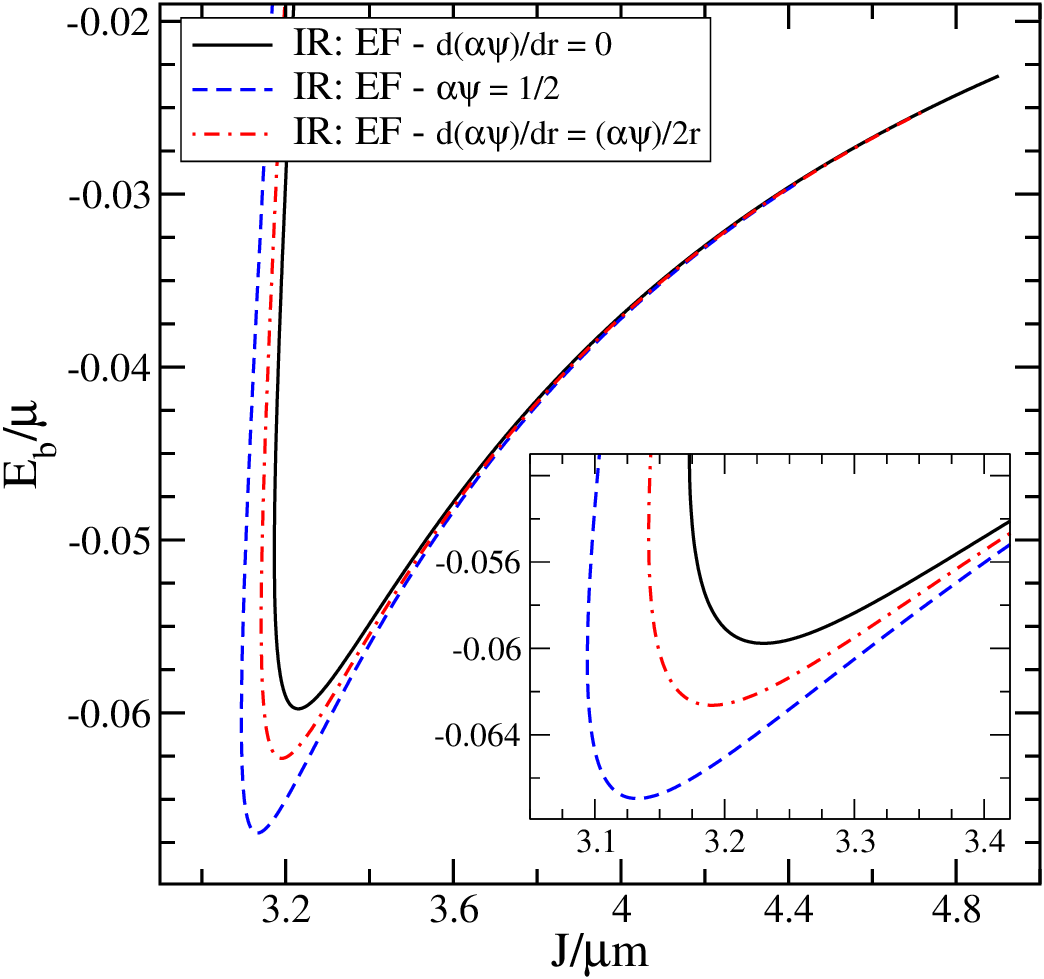}
\caption{Constant $M_{\rm irr}$ sequence of {\em irrotational} equal-mass
black holes.  Eddington--Finkelstein slicing is used in these cases,
and three different excision boundary conditions for the lapse were
used.}
\label{fig:Eb_J_EFIR}
\end{figure}

Figures~\ref{fig:Eb_J_EFCO} and \ref{fig:Eb_J_EFIR} display the
results for both corotating and irrotational sequences of equal-mass
black-hole binaries based on the Kerr--Schild-like slicing described
above.  In both cases, the same three lapse boundary conditions used
for the maximal slicing solutions were again used.  When the holes are
at large separation, the different lapse boundary conditions cause
little variation in the results.  However, when the holes are close
together, the different lapse boundary conditions cause significant
variation in the sequences.  From this example it seems that the
choice of the lapse boundary condition may have a significant effect
on QE solutions of the conformal thin-sandwich equations.  However,
this example may be somewhat misleading.

Maximal slicing ($\TrExCurv=0$) is based on a global geometric concept
that does not depend on the separation of the black holes in a binary.
For an isolated black hole, the Kerr--Schild slicing also has a
geometric interpretation.  However, a linear combination of the traces
of the extrinsic curvatures for individual black holes does not
retain this geometrical meaning.  Thus, as we
vary the separation of the black holes in the non-maximal slicing
sequences, we are also effectively varying the slicing condition.
This effect is weak when the holes are at large separation, but
becomes significant when the black holes are close together.

In constructing meaningful sequences, everything in the construction
of the individual models should be held fixed except for the
separation.  For maximal slicing, the various choices of the lapse
boundary condition choose a particular slice from among a family of
maximal slices.  However in the of the Kerr--Schild based slicing, is
seems likely that the functional form of $\TrExCurv$ as a function of
separation, and the form of the lapse boundary condition conspire to
define a different slicing condition for each initial-data model
considered.  Therefore, it seems reasonable to question the validity
of these non-maximal slicing sequences.  More importantly, we should
be cautious in attributing undue significance to the choice of the
lapse boundary condition based on this example.

\section{Discussion}
\label{sec:discussion}

In this paper, we have refined the QE boundary conditions defined
originally in Ref.~\cite{Cook-2002} and explored both single and
binary black-hole configurations.  The original motivation in deriving
these boundary conditions was to provide conditions that would be
consistent with quasiequilibrium configurations.  The binary
black hole initial-data sets constructed in Sec.~\ref{sec:quasi-circ-orbits}
are intended to be in quasiequilibrium.  While the individual black
holes in a binary cannot be in true equilibrium, it would be useful
to determine if they are roughly in equilibrium.

One measure of this is to see how well the QE lapse condition
(\ref{eq:QE_lapse_BC}) is satisfied.  For stationary black holes, this
equation holds.  However, in the general case, it does not.
Equation~(\ref{eq:QE_lapse_BC}) defines
$\Lapse_{BC}$ as the error in this boundary condition when applied on
the excision boundary of one of the holes.

\begin{figure}[!htbp]
\includegraphics[scale=0.465]{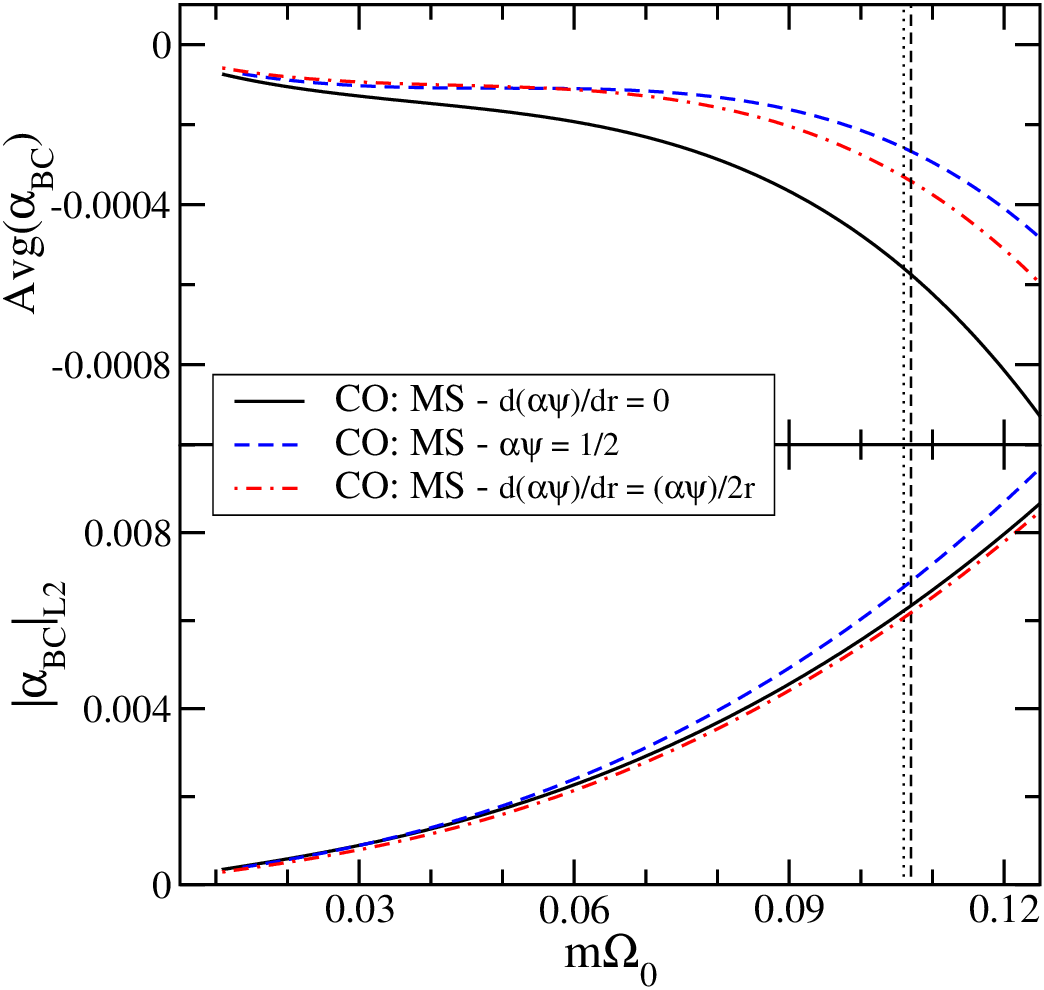}
\caption{Plot of the residual of the quasiequilibrium boundary
condition on the lapse (\ref{eq:QE_lapse_BC}) of one black hole.
Three different choices for the lapse boundary condition are shown for
corotating black holes in an equal-mass binary.  The upper plot shows
the average of the residual over the boundary surface.  The lower plot
shows the $L_2$-norm of the residual.  The vertical dashed line shows
the approximate location of the ISCO defined by a minimum in $E_b$.
The vertical dotted line shows the approximate location of the ISCO
defined by a minimum in $E_{\mbox{\tiny ADM}}$.}
\label{fig:QEBC__O_CO}
\end{figure}

\begin{figure}[!htbp]
\includegraphics[scale=0.465]{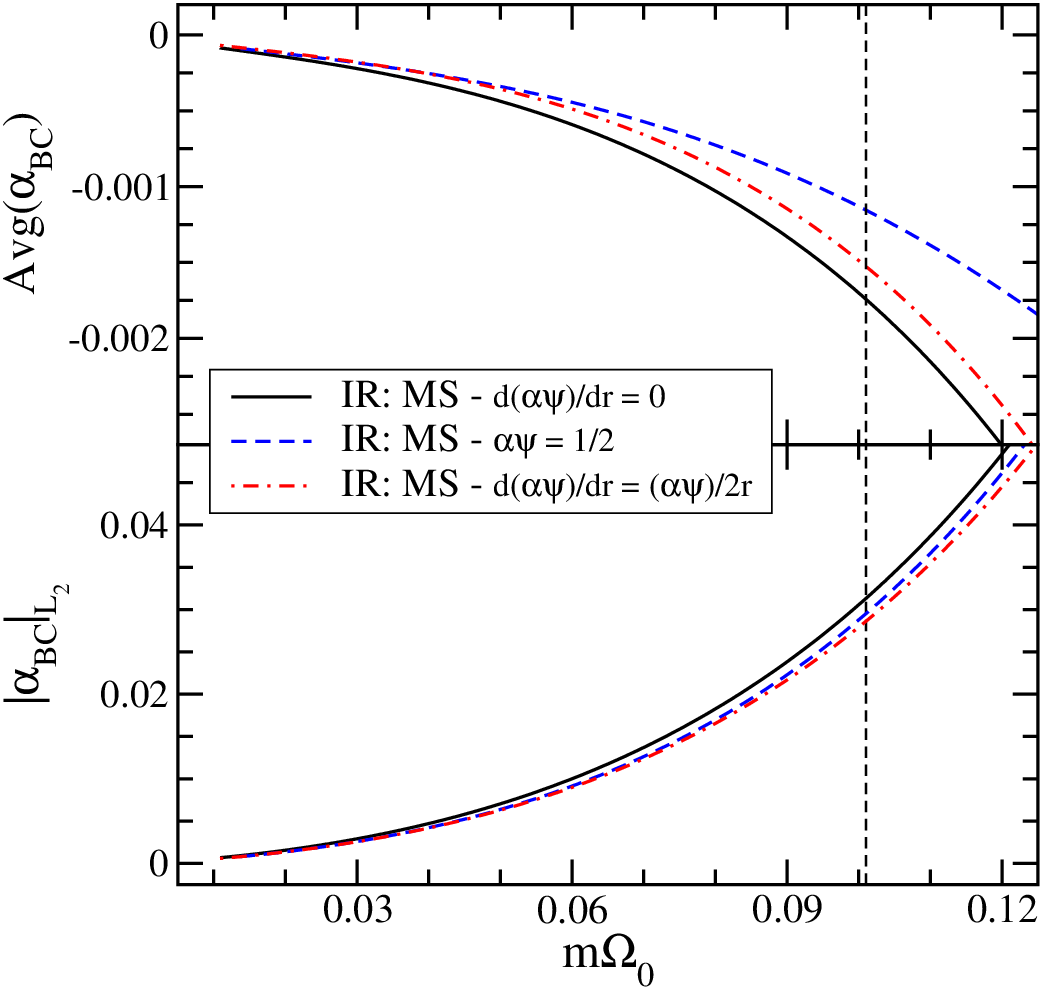}
\caption{Plot of the residual of the quasiequilibrium boundary
condition on the lapse (\ref{eq:QE_lapse_BC}) of one black hole.
Three different choices for the lapse boundary condition are shown for
irrotational black holes in an equal-mass binary.  The upper plot shows
the average of the residual over the boundary surface.  The lower plot
shows the $L_2$-norm of the residual.  The vertical dashed line shows
the approximate location of the ISCO defined by a minimum in $E_b$.}
\label{fig:QEBCL_O_IR}
\end{figure}

Figure~\ref{fig:QEBC__O_CO} shows both the average value of
$\Lapse_{BC}$ and the $L_2$ norm of $\Lapse_{BC}$ as a function of
$\Omega_0$ for a corotating equal-mass binary.
Figure~\ref{fig:QEBCL_O_IR} shows the same information for an
irrotational equal-mass binary.  At large separations (small
$\Omega_0$), we should expect that each black hole is nearly in
equilibrium.  For the corotating sequences, $m\Omega_0\approx0.01$
corresponds to a proper separation between the horizons of
approximately $20m$.  At this separation, $|\Lapse_{BC}|_{L_2} \approx
0.0003$.  At the ISCO separation, $m\Omega_0\approx0.11$ corresponding
to a proper separation between the horizons of approximately $4.5m$,
and $|\Lapse_{BC}|_{L_2}$ has increased by a factor of approximately
20.  As we might expect, the level of violation of the QE lapse
boundary condition increases steadily as the separation between
the holes decreases.  However, there is no dramatic increase in
the violation near the ISCO.  For the irrotational sequences,
$|\Lapse_{BC}|_{L_2}$ begins at $m\Omega_0\approx0.01$ at a level
approximately twice as large as that of the corotating sequence.
Near the ISCO, it has increased by a factor of approximately
50.

It seems that the rate of increase in the violation of the QE lapse
boundary condition for the irrotational sequences is faster than that 
seen in the corotating sequences.  This is not too surprising when
we recall that a true stationary binary configuration can only be
achieved for corotating binaries\cite{friedman-etal-2002}.  As with
the corotating sequences, the level of violation of the QE lapse 
boundary condition increases steadily as the separation between 
the holes decreases, and there is no dramatic increase in the
violation near the ISCO.

Another indicator of whether or not each black hole in the binary
is in equilibrium is given by the value of $\dtime\CF$ as evaluated
on the apparent horizon.  We can express the time derivative of the
conformal factor on any closed surface as
\begin{eqnarray}
\dtime\ln\CF &=& \mbox{$\frac14$}\Bigl[\CBCD_k\parShift^k 
+ 4\parShift^k\CBCD_k\ln\CF  
\label{eq:dt_psi}
\\ && \mbox{}\hspace{0.25in}
- \mbox{$\frac12$}\CBMetric_{k\ell}\CMtd^{k\ell}
+ \sqrt{2}\Oexpansion 
- \left(\Lapse - \perpShift\right)\BExCurv\Bigr].
\nonumber
\end{eqnarray}
Clearly, when the QE boundary conditions in
Eqs.~(\ref{eq:expansion_on_S}) and (\ref{eq:perpShift_BC}) are imposed on
the excision surface, the last two terms in Eq.~(\ref{eq:dt_psi})
vanish.  Furthermore, in constructing QE configurations, we have also
demanded that $\CMtd_{ij}=0$ globally.  Therefore, the only terms that
are possibly non-zero on the excision surface are those that involve
$\parShift^i$.

\begin{figure}[!htbp]
\includegraphics[scale=0.465]{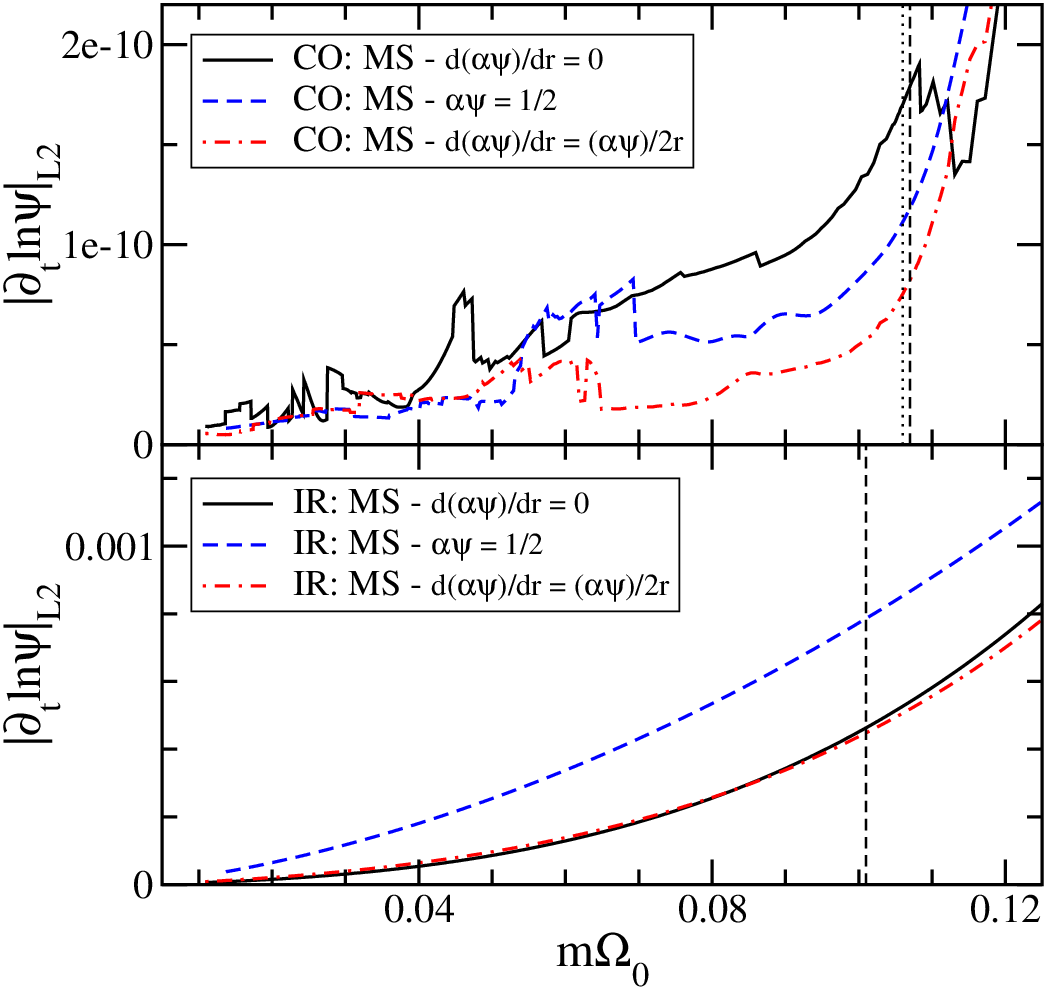}
\caption{Plot of the $L_2$-norm of $\dtime\ln\CF$ as evaluated on the
excision boundary of one black hole.  The upper half of the figure
shows the results for corotating black holes in an equal-mass binary,
while the lower half shows the results for irrotational black holes.
For both cases, three different choices for the lapse boundary
condition are shown.  In the upper half of the figure, the vertical
dashed line shows the approximate location of the ISCO defined by a
minimum in $E_b$.  The vertical dotted line shows the approximate
location of the ISCO defined by a minimum in $E_{\mbox{\tiny ADM}}$.  In the lower
half of the figure, the vertical dashed line shows the approximate
location of the ISCO defined by a minimum in $E_b$.}
\label{fig:Dtlnpsi_O}
\end{figure}

For corotating binaries, $\parShift^i=0$ and we find that the time
derivative of the conformal factor vanishes identically on the
excision surface.  This is confirmed in our numerical results as shown
in the upper half of Fig.~\ref{fig:Dtlnpsi_O}.  There, we see that
$\dtime\ln\CF=0$ to roundoff error.  For irrotational binaries, the QE
conditions require that we take $\parShift^i$ proportional to a
conformal Killing vector of $\CBMetric_{ij}$.  This implies that
$\parShift^i$ will also be a conformal Killing vector of
$\BMetric_{ij}$.  Unfortunately, the operator acting on $\parShift$
in Eq.~(\ref{eq:dt_psi}) is not the conformal Killing operator, but
rather
\begin{equation} 
\CBCD_k\parShift^k + 4\parShift^k\CBCD_k\ln\CF = \BCD_k\parShift^k.
\label{eq:dtpsi_shift_cond}
\end{equation}
While this would
vanish if $\parShift^k$ were a Killing vector of $\BMetric_{ij}$, it
will not vanish if $\parShift^k$ is only a conformal Killing vector of
$\BMetric_{ij}$, as it will be unless the configuration is truly
stationary.  Again, this is confirmed in our numerical results where
we find that $\dtime\ln\CF\sim10^{-5}$ when $m\Omega\sim0.01$ and
grows monotonically as the binary approaches the ISCO.  These results
are shown in the lower half of Fig.~\ref{fig:Dtlnpsi_O}.

The quasiequilibrium boundary conditions we have derived and tested
in this paper are extremely general.  Within the conformal
thin-sandwich approach, they will work for any number of black holes
that are to be considered in quasiequilibrium, or ``isolated''.  In
this paper, we have used several different choices for $\TrExCurv$, but
maintained the assumption of conformal flatness.  We emphasize that
this is not a limitation of the boundary conditions which can, in
fact, be used with any viable choice for the conformal three-geometry
specified by $\CMetric_{ij}$.  Furthermore, for binary systems, we
have only considered the special cases of corotating and irrotational
black holes.  Again, this is not a limitation of the boundary
conditions which can, in principle, produce any desired spin on the
individual black holes.

It has been pointed out that the boundary conditions we have derived
are precisely those required to construct a black hole satisfying the
isolated-horizon
conditions\cite{ashtekar_private_2002,ashtekar-etal-2000a,dreyer-etal-2003,ashtekar-krishnan-2003a}.
This is not surprising since the physical notions underlying an
isolated horizon and a black hole in quasiequilibrium are essentially
the same thing.  It seems likely that the unified approach offered by
the isolated-horizon framework will prove useful in further
understanding the physical content of the binary black hole initial
data constructed with the quasiequilibrium boundary conditions and to
further understand the role of the lapse boundary condition.  In fact,
during the final stages of the preparation of this manuscript, we
became aware of a paper by Jaramillo et al.\cite{Jaramillo-etal:2004}
that makes the connection between our quasiequilibrium boundary
conditions and isolated horizons more precise.  This paper argues that
the lapse boundary condition, Eq.~(\ref{eq:QE_lapse_BC}) derived
previously in Ref.~\cite{Cook-2002}, could be problematic, as we have
found and discussed, and shows that weakly isolated horizon
considerations do not restrict the lapse boundary condition when
constructing initial data, consistent with our findings here.  Furthermore,
they suggest an alternate boundary condition on the lapse, based on a
Lie-derivative along the null-generators of the horizon.  It is not
immediately clear that this proposed boundary condition can work.  For
the case of a static black hole, we have shown that essentially {\em
any} boundary condition on the lapse, when combined with the
quasiequilibrium boundary conditions, will yield a valid static slice
of Schwarzschild.  It may well be that this proposed boundary
condition is degenerate similar to Eq.~(\ref{eq:QE_lapse_BC}) (cf. our
discussion in Sec.~\ref{sec:bound-cond-lapse}).

Clearly, additional work is required to fully understand the boundary
conditions we have derived, and in particular the proper role of the
lapse boundary condition.  However, it is also clear that obtaining
appropriate boundary conditions is not the final issue in the quest to
construct astrophysically realistic binary black hole initial data.
The most pressing issue is the question of how to make a realistic
choice for the conformal three-geometry.  While the errors introduced
by the assumption of conformal flatness are not ``grave'', it is clear
that we must find a way to allow the physics to dictate the conformal
three-geometry instead of choosing it {\em a priori}.  The approach
along these directions outlined in Ref.~\cite{Shibata-etal-2004} (see
also Ref.~\cite{scheafer-gopakumar-2004}) is clearly promising.

\acknowledgments The authors are grateful to J.\ Isenberg, L.\
Lindblom, V.\ Moncrief, N.\ \`O~Murchadha, M.\ Scheel, M.\ Shibata,
and S.\ Teukolsky for illuminating discussions.  The authors are
grateful to K.\ Thorne and L.\ Lindblom for their hospitality during
the Caltech Visitors Program in the Numerical Simulation of
Gravitational Wave Sources in the spring of 2003, during which part of
this work was performed.  This work was supported in part by NSF
grants PHY-0140100 to Wake Forest University, PHY-9900672 to Cornell
University, and PHY-0244906 and PHY-0099568 to the California
Institute of Technology.  Computations were performed on the Wake
Forest University DEAC Cluster.

\appendix
\section{Corotating sequence}
In this Appendix, we list the numerical results for corotating
equal-mass black holes assuming conformal flatness, maximal slicing,
and using Eq.~(\ref{eq:Lapse-BC-2}) for the lapse boundary condition
on both excision surfaces.  The data has been scaled so that the
sequence satisfies Eq.~(\ref{eq:thermo_identity}) by following the
procedure outlined in
Eqs.~(\ref{eq:energy_scale})--(\ref{eq:thermo_scale}).  In order to
maintain accuracy in the scaling, the maximum coordinate separation
between successive models was $\Delta d=0.05$.  Data in the given
tables can be easily rescaled to construct sequences with $M_{\rm irr}$
held constant.

In Table~\ref{tab:corot_D_MS}, $d$ is the coordinate separation of the
centers of the excised regions.  $M_{\rm irr}$ is the irreducible mass
associated with {\em one} of the black holes.  $E_{\mbox{\tiny ADM}}$
is the ADM energy of the system.  $\Omega_0$ is the orbital angular
velocity of the binary system as measured at infinity.  $E_b$ is the
binding energy of the system defined as $E_b\equiv E_{\mbox{\tiny ADM}}
- 2M_{\rm irr}$.  $J_{\mbox{\tiny ADM}}$ is the total ADM angular momentum
of the binary system as measured at infinity.  Finally, $\ell$ is the
minimum proper separation between the two excision surfaces as measured
on the initial-data slice.

\begin{table}
\caption{Sequence of corotating equal-mass black holes on a maximal slice.
The length scale is set so that the ADM mass of the binary at infinite
separation is 1.  The ISCO is at separation $d=8.28$.}
\begin{ruledtabular}
\begin{tabular}{lllllll}
\multicolumn{1}{c}{$d$} & \multicolumn{1}{c}{$M_{\rm irr}$} 
& \multicolumn{1}{c}{$E_{\mbox{\tiny ADM}}-1$} & \multicolumn{1}{c}{$\Omega_0$} 
& \multicolumn{1}{c}{$E_b$} & \multicolumn{1}{c}{$J_{\mbox{\tiny ADM}}$} 
& \multicolumn{1}{c}{$\ell$} \\
\hline
40   & 0.5000000 & -0.0058296 & 0.01090 & -0.0058296 & 1.2280 & 21.81 \\
35   & 0.5000001 & -0.0065815 & 0.01327 & -0.0065816 & 1.1655 & 19.17 \\
30   & 0.5000003 & -0.0075478 & 0.01665 & -0.0075483 & 1.1005 & 16.52 \\
25   & 0.5000006 & -0.0088277 & 0.02175 & -0.0088289 & 1.0332 & 13.83 \\
20   & 0.5000013 & -0.0105789 & 0.03012 & -0.0105816 & 0.9647 & 11.11 \\
19   & 0.5000016 & -0.0110046 & 0.03246 & -0.0110079 & 0.9511 & 10.56 \\
18   & 0.5000020 & -0.0114600 & 0.03511 & -0.0114639 & 0.9376 & 10.00 \\
17   & 0.5000024 & -0.0119466 & 0.03814 & -0.0119514 & 0.9243 & 9.444 \\
16   & 0.5000030 & -0.0124654 & 0.04164 & -0.0124715 & 0.9113 & 8.882 \\
15   & 0.5000039 & -0.0130163 & 0.04571 & -0.0130240 & 0.8986 & 8.316 \\
14.5 & 0.5000044 & -0.0133032 & 0.04800 & -0.0133119 & 0.8925 & 8.031 \\
14   & 0.5000050 & -0.0135969 & 0.05049 & -0.0136068 & 0.8865 & 7.745 \\
13.5 & 0.5000059 & -0.0138966 & 0.05320 & -0.0139080 & 0.8807 & 7.458 \\
13   & 0.5000065 & -0.0142009 & 0.05617 & -0.0142140 & 0.8752 & 7.169 \\
12.5 & 0.5000076 & -0.0145079 & 0.05942 & -0.0145231 & 0.8699 & 6.879 \\
12   & 0.5000088 & -0.0148151 & 0.06300 & -0.0148328 & 0.8649 & 6.587 \\
11.5 & 0.5000104 & -0.0151187 & 0.06696 & -0.0151395 & 0.8602 & 6.293 \\
11   & 0.5000123 & -0.0154139 & 0.07135 & -0.0154385 & 0.8559 & 5.997 \\
10.5 & 0.5000147 & -0.0156939 & 0.07625 & -0.0157232 & 0.8521 & 5.699 \\
10   & 0.5000177 & -0.0159493 & 0.08174 & -0.0159847 & 0.8489 & 5.398 \\
9.5  & 0.5000217 & -0.0161675 & 0.08792 & -0.0162108 & 0.8463 & 5.095 \\
9    & 0.5000268 & -0.0163311 & 0.09492 & -0.0163846 & 0.8445 & 4.788 \\
8.9  & 0.5000280 & -0.0163553 & 0.09644 & -0.0164112 & 0.8442 & 4.726 \\
8.8  & 0.5000293 & -0.0163761 & 0.09799 & -0.0164346 & 0.8440 & 4.665 \\
8.7  & 0.5000306 & -0.0163934 & 0.09959 & -0.0164546 & 0.8438 & 4.603 \\
8.6  & 0.5000321 & -0.0164068 & 0.1012  & -0.0164709 & 0.8437 & 4.540 \\
8.5  & 0.5000336 & -0.0164160 & 0.1029  & -0.0164832 & 0.8436 & 4.478 \\
8.4  & 0.5000352 & -0.0164209 & 0.1047  & -0.0164913 & 0.8436 & 4.416 \\
8.35 & 0.5000361 & -0.0164215 & 0.1055  & -0.0164936 & 0.8436 & 4.384 \\
8.3  & 0.5000369 & -0.0164209 & 0.1064  & -0.0164948 & 0.8436 & 4.353 \\
8.28 & 0.5000373 & -0.0164203 & 0.1068  & -0.0164949 & 0.8436 & 4.341 \\
8.2  & 0.5000388 & -0.0164159 & 0.1083  & -0.0164934 & 0.8436 & 4.290 \\
8.1  & 0.5000407 & -0.0164054 & 0.1102  & -0.0164869 & 0.8437 & 4.227 \\
8    & 0.5000428 & -0.0163890 & 0.1121  & -0.0164747 & 0.8439 & 4.164
\end{tabular}
\end{ruledtabular}
\label{tab:corot_D_MS}
\end{table}

\section{Irrotational sequence}
In this Appendix, we list the numerical results for irrotational
equal-mass black holes assuming conformal flatness, maximal slicing,
and using Eq.~(\ref{eq:Lapse-BC-2}) for the lapse boundary condition
on both excision surfaces.  The data has been scaled so that the
sequence satisfies Eq.~(\ref{eq:thermo_identity}) by following the
procedure outlined in
Eqs.~(\ref{eq:energy_scale})--(\ref{eq:thermo_scale}).  In order to
maintain accuracy in the scaling, the maximum coordinate separation
between successive models was $\Delta d=0.05$.  Data in the given
tables can be easily rescaled to construct sequences with $M_{\rm irr}$
held constant.

In Table~\ref{tab:irrot_D_MS}, $d$ is the coordinate separation of the
centers of the excised regions.  $M_{\rm irr}$ is the irreducible mass
associated with {\em one} of the black holes.  $E_{\mbox{\tiny ADM}}$
is the ADM energy of the system.  $\Omega_0$ is the orbital angular
velocity of the binary system as measured at infinity.  $E_b$ is the
binding energy of the system defined as $E_b\equiv E_{\mbox{\tiny ADM}}
- 2M_{\rm irr}$.  $J_{\mbox{\tiny ADM}}$ is the total ADM angular momentum
of the binary system as measured at infinity.  Finally, $\ell$ is the
minimum proper separation between the two excision surfaces as measured
on the initial-data slice.

\begin{table}[!htbp]
\caption{Sequence of irrotational equal-mass black holes on a maximal slice.
The length scale is set so that the ADM mass of the binary at infinite
separation is 1.  The ISCO is at separation $d=8.69$.}
\begin{ruledtabular}
\begin{tabular}{lllllll}
\multicolumn{1}{c}{$d$} & \multicolumn{1}{c}{$M_{\rm irr}$} 
& \multicolumn{1}{c}{$E_{\mbox{\tiny ADM}}-1$} & \multicolumn{1}{c}{$\Omega_0$} 
& \multicolumn{1}{c}{$E_b$} & \multicolumn{1}{c}{$J_{\mbox{\tiny ADM}}$} 
& \multicolumn{1}{c}{$\ell$} \\
\hline
40   & 0.5000000 & -0.0058830 & 0.01090 & -0.0058830 & 1.2175 & 21.81 \\
35   & 0.4999988 & -0.0066618 & 0.01327 & -0.0066594 & 1.1527 & 19.18 \\
30   & 0.4999963 & -0.0076750 & 0.01665 & -0.0076676 & 1.0846 & 16.52 \\
25   & 0.4999903 & -0.0090444 & 0.02175 & -0.0090250 & 1.0127 & 13.84 \\
20   & 0.4999730 & -0.0109891 & 0.03014 & -0.0109351 & 0.9367 & 11.12 \\
19   & 0.4999661 & -0.0114788 & 0.03247 & -0.0114111 & 0.9211 & 10.57 \\
18   & 0.4999571 & -0.0120118 & 0.03513 & -0.0119261 & 0.9053 & 10.01 \\
17   & 0.4999451 & -0.0125938 & 0.03817 & -0.0124839 & 0.8894 & 9.454 \\
16   & 0.4999286 & -0.0132310 & 0.04169 & -0.0130882 & 0.8734 & 8.893 \\
15   & 0.4999057 & -0.0139306 & 0.04578 & -0.0137419 & 0.8574 & 8.329 \\
14.5 & 0.4998908 & -0.0143062 & 0.04808 & -0.0140878 & 0.8494 & 8.045 \\
14   & 0.4998729 & -0.0147003 & 0.05059 & -0.0144462 & 0.8414 & 7.759 \\
13.5 & 0.4998513 & -0.0151140 & 0.05332 & -0.0148166 & 0.8334 & 7.473 \\
13   & 0.4998248 & -0.0155484 & 0.05632 & -0.0151980 & 0.8255 & 7.185 \\
12.5 & 0.4997922 & -0.0160042 & 0.05961 & -0.0155887 & 0.8176 & 6.895 \\
12   & 0.4997517 & -0.0164824 & 0.06324 & -0.0159857 & 0.8098 & 6.604 \\
11.5 & 0.4997006 & -0.0169834 & 0.06726 & -0.0163846 & 0.8022 & 6.311 \\
11   & 0.4996356 & -0.0175072 & 0.07174 & -0.0167784 & 0.7946 & 6.016 \\
10.5 & 0.4995517 & -0.0180530 & 0.07676 & -0.0171564 & 0.7873 & 5.718 \\
10   & 0.4994416 & -0.0186189 & 0.08241 & -0.0175020 & 0.7802 & 5.419 \\
9.5  & 0.4992944 & -0.0192006 & 0.08882 & -0.0177893 & 0.7733 & 5.116 \\
9    & 0.4990934 & -0.0197901 & 0.09616 & -0.0179770 & 0.7670 & 4.811 \\
8.9  & 0.4990447 & -0.0199079 & 0.09776 & -0.0179972 & 0.7658 & 4.749 \\
8.8  & 0.4989924 & -0.0200252 & 0.09940 & -0.0180101 & 0.7646 & 4.688 \\
8.69 & 0.4989306 & -0.0201537 & 0.1013  & -0.0180149 & 0.7633 & 4.620 \\
8.6  & 0.4988763 & -0.0202582 & 0.1029  & -0.0180108 & 0.7623 & 4.564 \\
8.5  & 0.4988117 & -0.0203733 & 0.1047  & -0.0179966 & 0.7611 & 4.502 \\
8    & 0.4984034 & -0.0209241 & 0.1146  & -0.0177308 & 0.7561 & 4.189 \\
7.5  & 0.4977854 & -0.0213930 & 0.1265  & -0.0169638 & 0.7522 & 3.872 \\
7    & 0.4967902 & -0.0216779 & 0.1411  & -0.0152582 & 0.7500 & 3.548 \\
6.9  & 0.4965184 & -0.0216959 & 0.1445  & -0.0147327 & 0.7499 & 3.483 \\
6.86 & 0.4964004 & -0.0216980 & 0.1459  & -0.0144988 & 0.7499 & 3.457 \\
6.7  & 0.4958660 & -0.0216708 & 0.1518  & -0.0134029 & 0.7501 & 3.351 \\
6.5  & 0.4950163 & -0.0215319 & 0.1600  & -0.0115644 & 0.7510 & 3.218
\end{tabular}
\end{ruledtabular}
\label{tab:irrot_D_MS}
\end{table}

\bibliography{}

\end{document}